\pdfoutput=1

\documentclass[3p,times]{elsarticle}

\usepackage[pdftex]{hyperref}

\usepackage{ecrc-arXiv}

\usepackage{color}

\volume{}

\firstpage{1}

\journalname{}

\runauth{G. Konrad et al.}

\jid{}

\jnltitlelogo{}

\usepackage{amssymb}

\usepackage[figuresright]{rotating}

\newcommand{\aspect}{\textit{a}SPECT}

\begin{document}

\begin{frontmatter}



\dochead{}

\title{The magnetic shielding for the neutron decay spectrometer \aspect}


\author[1,2]{Gertrud Konrad\corref{cor1}}
\cortext[cor1]{Corresponding author. Present address: Technische Universit\"at Wien, Atominstitut, Stadionallee 2, A-1020 Wien, Austria.}
\ead{gkonrad@ati.ac.at}
\author[1]{Fidel Ayala Guardia}
\author[1,3]{Stefan Bae{\ss}ler}
\author[1]{Michael Borg}
\author[4,5]{Ferenc Gl\"uck}
\author[1]{Werner Heil}
\author[1,6]{Stefan Hiebel}
\author[1]{Raquel Mu\~noz Horta}
\author[1]{Yury Sobolev}

\address[1]{Universit\"at Mainz, Institut f\"ur Physik, Staudingerweg 7, D-55128 Mainz, Germany}
\address[2]{Technische Universit\"at Wien, Atominstitut, Stadionallee 2, A-1020 Wien, Austria}
\address[3]{University of Virginia, Department of Physics, Charlottesville, VA 22904, USA}
\address[4]{Institute for Nuclear Physics (IKP), Karlsruhe Institute of Technology, P.O.B. 3640, D-76021 Karlsruhe, Germany}
\address[5]{Wigner Research Institute for Physics, P.O.B. 49, H-1525 Budapest, Hungary}
\address[6]{Sekels GmbH, Dieselstra{\ss}e 6, D-61239 Ober-M\"orlen, Germany}

\begin{abstract}
Many experiments in nuclear and neutron physics are confronted with the problem that they use a superconducting magnetic spectrometer which potentially affects other experiments by their stray magnetic field.  The retardation spectrometer {\aspect} consists, inter alia, of a superconducting magnet system that produces a strong longitudinal magnetic field of up to 6.2\,T.  In order not to disturb other experiments in the vicinity of {\aspect}, we had to develop a magnetic field return yoke for the magnet system.  While the return yoke must reduce the stray magnetic field, the internal magnetic field and its homogeneity should not be affected.  As in many cases, the magnetic shielding for {\aspect} must manage with limited space.  In addition, we must ensure that the additional magnetic forces on the magnet coils are not destructive.

In order to determine the most suitable geometry for the magnetic shielding for {\aspect}, we simulated a variety of possible geometries and combinations of shielding materials of non-linear permeability.  The results of our simulations were checked through magnetic field measurements both with Hall and nuclear magnetic resonance probes.  The experimental data are in good agreement with the simulated values:  The mean deviation from the simulated exterior magnetic field is $(-1.7 \pm 4.8)$\,\%.  However, in the two critical regions, the internal magnetic field deviates by $0.2$\,\% respectively $<1 \times 10^{-4}$ from the simulated values.
\end{abstract}


\end{frontmatter}



\section{Introduction}
\label{sec:intro}

The neutron decay spectrometer {\aspect} \cite{zimmer:2000, glueck:2005, baessler:2008}, introduced in Sec.~\ref{sec:aspect}, has been built to measure the antineutrino-electron angular correlation coefficient $a$ with unprecedented precision.  The spectrometer consists, inter alia, of a superconducting magnet system that generates a stray magnetic field of 5\,Gauss in a radial distance of 5\,m, cf. Fig.~\ref{fig:RadialField}.

The {\aspect} experiment had to move to the cold neutron beam line PF1b \cite{PF1} of the Institut Laue-Langevin (ILL) in Grenoble, France \cite{konrad:2009,simson:2009}.  In order not to disturb neighboring experiments, in particular the spin-echo spectrometer IN11 \cite{IN11}, by the stray magnetic field, we had to develop a magnetic field return yoke \cite{konrad:2007,konrad:2011c} for the {\aspect} magnet system.  The stray magnetic field must therefore be suppressed to less than 1\,Gauss (0.1\,mT) in a radial distance of 5\,m.  This corresponds to a reduction in stray field by a factor of about 10.

\begin{figure}[t]%
\centering
\includegraphics[width=0.93\textwidth]{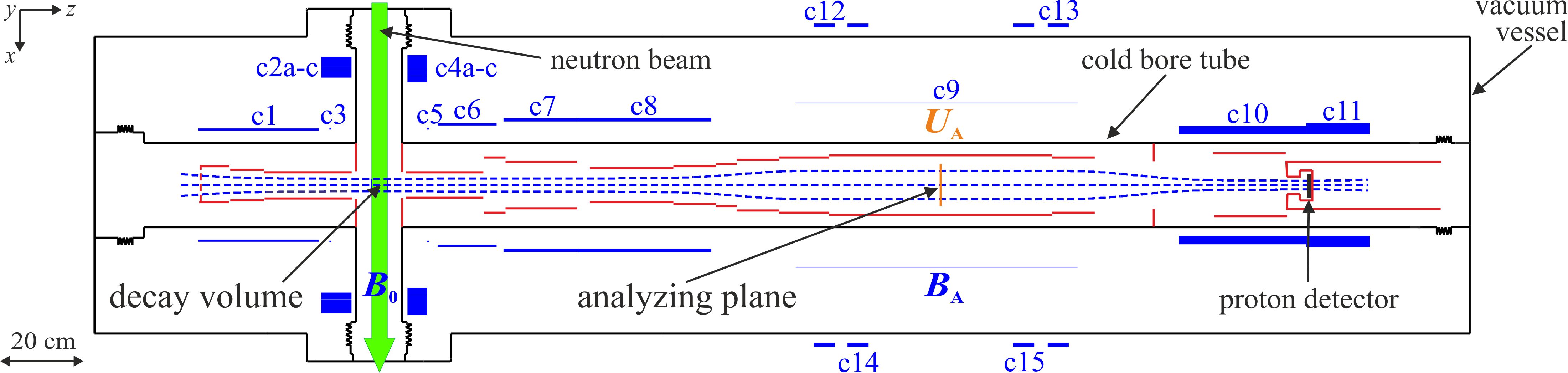}%
\caption{Sketch of the electromagnetic set-up of the {\aspect} experiment at the Institut Laue-Langevin in Grenoble, France.  The magnet coils are drawn in blue and are denoted as c1 to c15.  The electrodes are drawn in red.  Cold, unpolarized neutrons (green) pass through the decay volume where only a few neutrons decay.  The decay protons are guided by the strong magnetic field (dashed blue) towards the analyzing plane (orange) and subsequently the proton detector (black).  Protons emitted in the negative $z$-direction are reflected by the electrostatic mirror (dashed red).  The analyzing plane voltage $U_{\rm A}$ is varied to scan the shape of the proton recoil spectrum.  We note that the sketch is rotated by 90$^\circ$.}%
\label{fig:Sketch}%
\end{figure}

Magnets can be shielded either actively or passively.  An active shielding consists of secondary shielding coils surrounding the primary ones.  The shielding coils are designed to produce a magnetic field that reduces the stray magnetic field of the primary coils.  However, active shielding works very well only if the diameter of the secondary coils is substantially larger than that of the primary ones.  A ratio of the radii of the secondary to the primary coils of 3 to 2 is reasonable.  On the other hand, the free space at the PF1b beam position is limited to about $9.5\times2.3$\,m$^2$ ($\mbox{length}\times\mbox{width}$) and $^{+4.1}_{-1.4}$\,m in height (relative to the center of the decay volume of {\aspect} shown in Fig.~\ref{fig:Sketch}).  In the case of the {\aspect} primary coils, with coil diameters of up to 60\,cm as indicated in Table~\ref{tab:coils} (see also Fig.~\ref{fig:Sketch}), active shielding would manage with the limited space conditions.  But due to price (and chronological order of events), active shielding is no solution for the {\aspect} magnet system.

A passive shielding consists of high magnetic permeability materials which enclose the magnet system.  The best geometry for passive shielding is a closed container surrounding the magnet coils, as considered in Sec.~\ref{sec:axially}.  Soft magnetic materials, i.e., ferromagnetic materials that can be easily magnetized at low magnetic field, have a high magnetic permeability.  However, the magnetization of the shielding material causes a substantial magnetic field within, as illustrated below in Figs.~\ref{fig:DVshielding} and~\ref{fig:APshielding}, and in the interior of the magnet coils.  The former requires that the additional magnetic field produced by the shielding material neither disturbs the shape of the magnetic field of {\aspect} nor its high homogeneity in the order of $1\times10^{-4}$ both in the decay volume (DV) and in the analyzing plane (AP, shown in Fig.~\ref{fig:Sketch}), cf. Figs.~\ref{fig:DVshielding} and~\ref{fig:APshielding}.  The latter causes additional magnetic forces on the {\aspect} coils.  The manufacturer of the {\aspect} magnet system, Cryogenics Ltd., recommends that the magnetic forces on the magnet coils do not change their sign relative to their original design which did not have a passive shielding, and that the relative force changes are small.  

According to Earnshaw's theorem \cite{earnshaw:1842}, stable equilibrium corresponds to minimum potential energy, whereas unstable equilibrium corresponds to maximum potential energy.  Figure~\ref{fig:photo} shows the spectrometer {\aspect} within its magnetic field return yoke, in its equilibrium position.  In order to keep the magnet in position, and to resist its natural tendency to decrease its (gravitational) potential energy by falling down, we need to attach it to the return yoke.  In addition, a displacement of the magnet towards, e.g., one of the billets results in a decrease in (magnetic) potential energy.  In order to avoid that the magnet would be attracted to this billet and consequently be damaged by strong magnetic forces, we need to align the spectrometer within its return yoke.  

To sum up, the design of the magnetic shielding for the neutron decay spectrometer {\aspect} had to take into consideration the following requirements:
\begin{enumerate}
	\item The stray magnetic field must be suppressed to less than 1\,Gauss in a radial distance of 5\,m from the DV.\label{i:R1}
	\item The magnetic shielding must manage with the limited space conditions at the beam position.\label{i:R2}
	\item The relevant properties of the internal magnetic field, mainly its homogeneity, should not be disturbed.\label{i:R3}
	\item The additional magnetic forces on the magnet coils must not be destructive.\label{i:R4}
  \item The magnetic shielding should not prevent access to the spectrometer.
	\item The spectrometer must be aligned within its magnetic field return yoke.\label{i:R5}
\end{enumerate}

Requirement \ref{i:R1} needs lots of shielding material, requirements \ref{i:R3} and \ref{i:R4} want the magnetic shielding far away from the magnet coils, but requirement \ref{i:R2} and price work in the opposite direction.

Previous, competing, and future experiments in neutron beta decay were or are faced with a similar problem:  For the electron spectrometer PERKEO II \cite{abele:1997,abele:2002,kreuz:2005a,schumann:2007c,schumann:2008b,mund:2013} and its successor PERKEO III \cite{maerkisch:2006,maerkisch:2009,mest:2011} steel shieldings were installed in order not to disturb the spin-echo spectrometer IN11 at the ILL.  The magnet coils of the aCORN experiment \cite{wietfeldt:2005b,wietfeldt:2009} are surrounded with an iron structure that acts as a support for aCORN as well as a flux return.  The superconducting magnet coil system of the funded Nab experiment \cite{bowman:2005,pocanic:2009,alarcon:2010,baessler:2012} will be actively shielded to comply with the SNS (Spallation Neutron Source in Oak Ridge, Tennesse) policy about stray magnetic fields.  In addition, there will be a passive magnetic field return yoke made of steel, similar to the one described in Sec.~\ref{sec:nonaxially}.  The new facility PERC \cite{dubbers:2008,konrad:2012} will be passively shielded to comply with the safety regulations at the Forschungs-Neutronenquelle Heinz Maier-Leibnitz (FRM II) in Munich, Germany.  Its superconducting magnet system will be surrounded with a soft iron / construction steel structure \cite{haiden:2013,haiden:2014} that expands the return yoke described in Sec.~\ref{sec:nonaxially} by steel plates as used for the PERKEO III experiment.  

There are also experiments in nuclear beta decay that are confronted with a similar problem:  The WITCH \cite{beck:2003,beck:2011a} magnets cause an interfering stray magnetic field in the beam lines of the REX-ISOLDE facility \cite{habs:2000,REXISOLDE}.  It was therefore decided to install a mumetal magnetic shield around the REX-ISOLDE set-up \cite{tandecki:2011}.

For the {\aspect} magnet system, we have considered a frame consisting of a bottom and a top plate and 4 billets as the most suitable geometry for the magnetic field return yoke.  The design of the return yoke is introduced in Sec.~\ref{sec:nonaxially}.  The numerical integration of the complete elliptical integrals \cite{garrett:1963}, the TOSCA analysis package \cite{TOSCA}, and the finite element analysis (FEA) software package COMSOL Multiphysics$^{\textregistered}$ \cite{COMSOL} have been used to model and optimize the design.  Experimental verification and validation of the finite element model are described in Sec.~\ref{sec:results}.

\section{The spectrometer \aspect}
\label{sec:aspect}

{\aspect} is a retardation spectrometer which measures the proton recoil spectrum in the decay of free neutrons by counting all decay protons that overcome an electrostatic barrier \cite{zimmer:2000,glueck:2005}.  From the proton spectrum the antineutrino-electron angular correlation coefficient $a$ can be derived.  The electromagnetic set-up of {\aspect} is shown in Fig.~\ref{fig:Sketch} and in Ref.~\cite[Sec.~2]{glueck:2005}.  Its design principles and systematics are thoroughly discussed in Ref.~\cite{glueck:2005}.

The superconducting magnet system of the {\aspect} experiment consists of 11 coils, denoted as c1 to c11, placed inside a cylinder with a length of 3.2\,m and a diameter of 70\,cm.  Details on the coils can be found in Table~\ref{tab:coils}.  The other coils shown give small corrections\footnote{After use of the correction coils c3, c5, c12 and c13, the magnetic field of {\aspect} met its specifications \cite{ayala:2005}.  An additional pair of (external) correction coils, namely c14 and c15, serves to change the magnetic field in the AP by up to 1\,\%, for the investigation of systematic effects \cite{konrad:2009,simson:2009}.}, and are not relevant to the further discussion in this paper.

\begin{figure}[htbp]%
  \begin{minipage}[t]{0.475\textwidth}
    \includegraphics[width=0.96\textwidth]{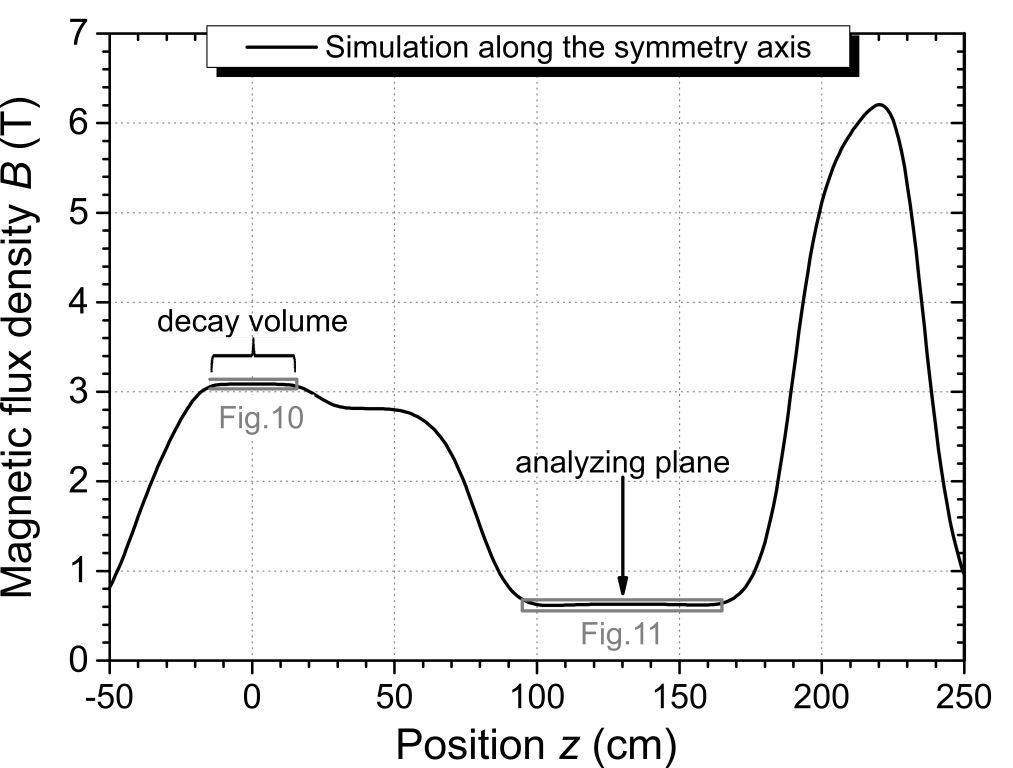}%
    \caption{Magnetic flux density $B(z)$ along the symmetry axis of the spectrometer {\aspect}, for its design current of $I_{\rm main}=100$\,A.  The enlargements (gray boxes) of the decay volume and of the analyzing plane are shown below in Figs.~\ref{fig:DVshielding} and~\ref{fig:APshielding}, respectively.}%
    \label{fig:LongitudinalField}%
  \end{minipage}
  \hfill
  \begin{minipage}[t]{0.475\textwidth}
		\includegraphics[width=\textwidth]{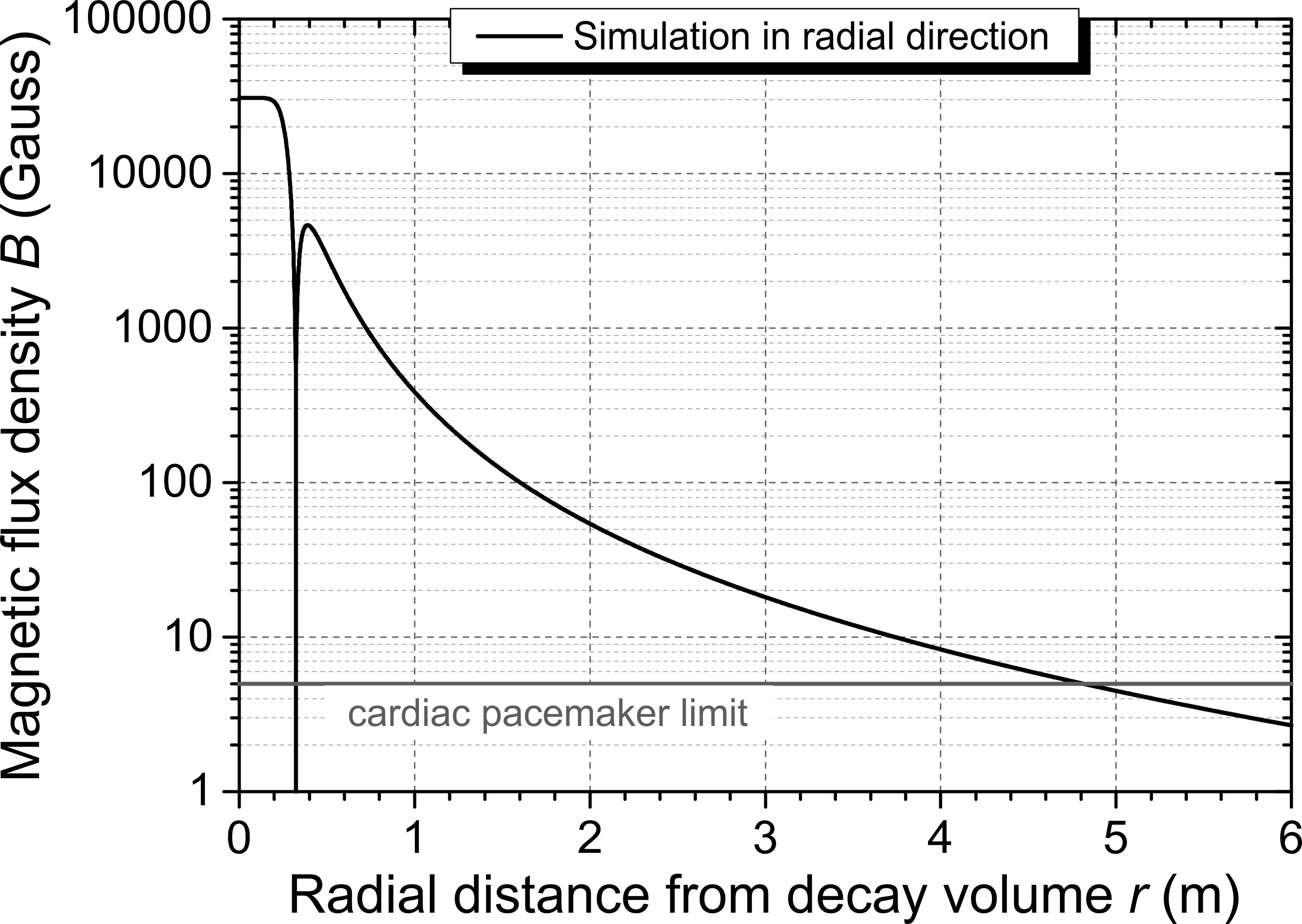}%
		\caption{Magnetic flux density $B(r)$ in radial direction from the decay volume (DV) of the spectrometer \aspect, for its design current of $I_{\rm main}=100$\,A.  The cardiac pacemaker limit (gray line) of 5\,Gauss is reached at a radial distance of 4.8\,m from the DV.}%
		\label{fig:RadialField}%
  \end{minipage}
\end{figure}

The (superconducting) magnet coil system, schematically shown in Fig.~\ref{fig:Sketch}, and its magnetic field are axially symmetric.  Figures~\ref{fig:LongitudinalField} and~\ref{fig:RadialField} show that the magnetic field varies from 0.6\,T to 6.2\,T along its symmetry axis and down to the cardiac pacemaker limit of 5\,Gauss (0.5\,mT) in a radial distance of 4.8\,m from the DV, respectively, both at the design current of $I_{\rm main}=100$\,A.

The shape of the magnetic field both in the DV and in the AP is critical for the operation of the {\aspect} spectrometer.  In particular, the ratio of the magnetic fields in the AP and in the DV, $r_{\rm B} = B_{\rm A} / B_0$, enters into the determination of the angular correlation coefficient $a$ and has to be known precisely.  The enlargements of the DV and of the AP, shown in Figs.~\ref{fig:DVshielding} and~\ref{fig:APshielding} below, respectively, illustrate the high homogeneity of the magnetic field which enables a measurement with a relative accuracy of the magnetic field ratio $r_{\rm B}$ in the order of $1\times10^{-4}$ (see also Ref.~\cite{ayala:2011}).

\begin{table}[htbp]%
\caption{Coil geometry of the {\aspect} magnet system.  The first column indicates the description of the coil.  The second column specifies the coil type using the following abbreviations: n (normal conducting) and s (superconducting).  The third to sixth column list the inner radius, the radial thickness, the lower edge, and the length of the coil.  The last column gives the current density of the coil, in units of the supply current in amperes.}
\centering
\begin{tabular}{l|c|rrrr|r}
  coil & type & $r_{\rm in}$ (cm) & $t$ (cm) & $z_{\rm min}$ (cm) & $l$ (cm) & $J_\varphi$ (A\,cm$^{-2}$) \\
  \hline
  c1 & s & 12.90 & 0.48\hphantom{0} & -42.94 & 28.66\hphantom{0} & 256.53 $\times I_{\rm main}$\\
  c2a & s & 25.48 & 1.60\hphantom{0} & -13.70 & 7.15\hphantom{0} & 146.85 $\times I_{\rm main}$\\
  c2b & s & 27.08 & 0.70\hphantom{0} & -13.70 & 7.15\hphantom{0} & 189.81 $\times I_{\rm main}$\\
  c2c & s & 27.78 & 2.622 & -13.70 & 7.15\hphantom{0} & 256.46 $\times I_{\rm main}$\\
  c4a & s & 24.28 & 1.76\hphantom{0} & 6.73 & 4.68\hphantom{0} & 146.90 $\times I_{\rm main}$\\
  c4b & s & 26.04 & 1.26\hphantom{0} & 6.73 & 4.68\hphantom{0} & 189.26 $\times I_{\rm main}$\\
  c4c & s & 27.30 & 3.48\hphantom{0} & 6.73 & 4.68\hphantom{0} & 256.41 $\times I_{\rm main}$\\
  c6 & s & 14.08 & 0.48\hphantom{0} & 13.96 & 13.97\hphantom{0} & 256.50 $\times I_{\rm main}$\\
  c7 & s & 15.00 & 0.78\hphantom{0} & 29.54 & 17.81\hphantom{0} & 256.41 $\times I_{\rm main}$\\
  c8 & s & 15.00 & 0.90\hphantom{0} & 47.41 & 31.59\hphantom{0} & 256.41 $\times I_{\rm main}$\\
  c9 & s & 19.28 & 0.18\hphantom{0} & 99.06 & 66.89\hphantom{0} & 256.39 $\times I_{\rm main}$\\
  c10 & s & 12.00 & 1.98\hphantom{0} & 190.00 & 30.29\hphantom{0} & 256.41 $\times I_{\rm main}$\\
  c11 & s & 12.00 & 2.70\hphantom{0} & 220.29 & 15.02\hphantom{0} & 256.32 $\times I_{\rm main}$\\
  \hline
  c3 & s & 13.13 & 0.42\hphantom{0} & -11.78 & 0.455 & 256.41 $\times I_3$\hphantom{00}\\
  c5 & s & 13.13 & 0.42\hphantom{0} & 11.32 & 0.455 & 256.41 $\times I_5$\hphantom{00}\\
	\hline
  c12 & n & 37.40 & 2.10\hphantom{0} & 102.00 & 5.00\hphantom{0} & 3.05 $\times I_{\rm ahc}$\hphantom{0}\\
  c13 & n & 37.40 & 2.10\hphantom{0} & 156.50 & 5.00\hphantom{0} & -3.05 $\times I_{\rm ahc}$\hphantom{0}\\
  \hline
  c14 & n & 37.40 & 2.80\hphantom{0} & 109.50 & 5.00\hphantom{0} & 2.86 $\times I_{\rm hc}$\hphantom{0\,}\\
  c15 & n & 37.40 & 2.80\hphantom{0} & 148.50 & 5.00\hphantom{0} & 2.86 $\times I_{\rm hc}$\hphantom{0\,}\\
\end{tabular}
\label{tab:coils}
\end{table}

\section{2D axially symmetric shielding}
\label{sec:axially}

We have started to design a passive shielding with simulations in two dimensions (\textit{Axial symmetry}), since the {\aspect} coil system described in Sec.~\ref{sec:aspect} is axially symmetric.

The magnetic flux density $\textbf{B}$ of a finite-length solenoid aligned with the $z-$axis, centered around zero, with infinitely thin walls, radius $R$, and length $L$ is given by \cite{garrett:1963}
\begin{eqnarray}
  \textbf{B} & = & B_\rho \hat{\textbf{e}}_\rho + B_{\rm z} \hat{\textbf{e}}_{\rm z}  
	\label{eq:Baxi}  \\
  B_\rho & = & \frac{\mu_0}{\pi} \frac{N I}{L} \sqrt{\frac{R}{\rho}}
							 \left[ \frac{k^2-2}{2k} K(k) + \frac{1}{k} E(k)
							 \right]^{z+L/2}_{z-L/2}  \nonumber  \\
  B_{\rm z} & = & \frac{\mu_0}{4 \pi}  \frac{N I}{L} \frac{1}{\sqrt{R \rho}}
							 \left[ \zeta k \left( K(k) + \frac{R-\rho}{R+\rho} \Pi(h,k) \right)
							 \right]^{z+L/2}_{z-L/2}  \nonumber 
\end{eqnarray}
where $\hat{\textbf{e}}_\rho$ is the (cylindrical) radial unit vector, $\hat{\textbf{e}}_{\rm z}$ the Cartesian unit vector codirectional with the $z-$axis, $\mu_0=4\pi\times10^{-7}$\,H\,m$^{-1}$ the permeability of vacuum, $N$ the number of turns, $I$ the current through the solenoid, flowing counter clockwise as seen from above ($z>0$), $\rho$ and $z$ are the cylindrical coordinates of the point where the magnetic field is calculated, and $K(k)$, $E(k)$, and $\Pi(h,k)$ are the complete elliptic integral functions of the first, second, and third kind, respectively, as functions of
\begin{eqnarray}
  h &=& \frac{2 \sqrt{R \rho}}{R+\rho}  \nonumber  \\
	k &=& \frac{2 \sqrt{R \rho}}{\sqrt{(R+\rho)^2+\zeta^2}}  \nonumber
\label{eq:Bpara}
\end{eqnarray}
Then the magnetic flux density of a thick, finite-length solenoid is obtained by numerical integration over the solenoid's thickness $t$.

In addition, the magnetization $\textbf{M}_{\rm shield}$ in a shielding material is defined by
\begin{eqnarray}
  \textbf{M}_{\rm shield} & = & \chi_{\rm m} \textbf{H}_{\rm in},
\label{eq:Mshield}
\end{eqnarray}
where $\chi_{\rm m}=\mu_{\rm r}-1$ and $\mu_{\rm r}$ are the susceptibility and the relative permeability of the shielding material, respectively, taken as constant, and $\textbf{H}_{\rm in}$ is the magnetic field strength in the shielding material.  The internal magnetic field $\textbf{H}_{\rm in}$ is composed of the magnetizing fields $\textbf{B}_i$ generated externally by the {\aspect} coils c$i$ ($i=1, 2a,\ldots, 11$) and the demagnetizing field $\textbf{H}_{\rm d}$ generated by the magnetization of the shielding material:
\begin{eqnarray}
  \textbf{H}_{\rm in} &=& \frac{1}{\mu_0} \sum_{i=1}^{11} \textbf{B}_i + \textbf{H}_{\rm d}
\label{eq:Hin}
\end{eqnarray}
with $\textbf{B}_i$ from Eq.~(\ref{eq:Baxi}).  Thus, we can calculate the magnetic field $\textbf{B}_{\rm shield}$ caused by the shielding at any point $\textbf{r}$ in space in the dipole approximation:
\begin{eqnarray}
  \textbf{B}_{\rm shield}(\textbf{r}) & = & \frac{\mu_0}{4\pi} 
		\int\limits_{V_{\rm shield}} {\rm d}^3 r^\prime 
		\left( \frac{3\textbf{r} \left(\textbf{M}_{\rm shield}(\textbf{r}^\prime)\cdot\textbf{r} \right)}{r^5} - \frac{\textbf{M}_{\rm shield}(\textbf{r}^\prime)}{r^3} \right),
\label{eq:Bshield}
\end{eqnarray}
where $V_{\rm shield}$ is the volume of the shielding and $r=|\textbf{r}|$.

For the 2D axially symmetric shielding calculations, the {\aspect} coil system was surrounded with a can, i.e. a cylinder with a cap on top and on bottom.  For the purpose of the calculation of the magnetic field strength $\textbf{H}$ of the shielding material, we have assumed that both the inner and the outer surface of the cylinder and of the caps are covered by a fictitious magnetic surface charge.  In this way, we can use a boundary element formulation which is similar to the calculation of an electric field generated by a surface charge density distribution.  According to Eq.~(3) in Ref.~\cite{babic:2000}, the magnetic surface charge density distribution of the shielding material is determined from an integral equation using the magnetic field generated by the {\aspect} coils and the geometry and the constant permeability of the shielding material.  In order to solve the integral equation, the surface of the shielding is discretized by a few hundred cylindrical and disk-shaped elements, with constant magnetic charge density over the surface of each element.  The magnetic field of these elements is obtained by numerical integration of the complete elliptic integrals, similarly to the calculation of the electric field of axially symmetric electrodes.  Then the discretization leads to a linear algebraic equation system for the unknown magnetic charge densities, which is solved using Gauss-Jordan elimination.  The magnetic field outside the shielding is derived as the superposition of the field of the {\aspect} coils and the field $\textbf{H}_{\rm d}$ of the shielding material using the calculated values of the magnetic charge density.  In order to test our numerical calculations, we calculated with our code the magnetic field of a magnetized sphere and the magnetic shielding of a hollow sphere, and compared our results with the corresponding analytical formulas.  In both cases, complete agreement between numerical and analytical results were found.

For the simulation of shielding materials of non-constant permeability, and geometries without axial symmetry (presented in Sec.~\ref{sec:nonaxially}), we had to switch to a finite element analysis software (FEA).  The finite element method (FEM) is a numerical method for determining approximate solutions to partial differential equations (PDE).  In simple terms, FEM is a method for dividing a domain into a finite number of small elements, reconnected at nodes in which certain properties are assumed to be constant which in reality are not.  For each element, the field quantity is interpolated by a polynomial, what results in a set of simultaneous algebraic equations.  To model and optimize the design of a passive shielding for {\aspect}, the FEA software package COMSOL Multiphysics$^{\textregistered}$, version 3.2b, Electromagnetics Module\footnote{The Electromagnetics Module ceased to be developed after version 3.2b.  It was replaced and further developed by the AC/DC and RF Modules, with similar or enhanced functionality.}, has been used.

\begin{figure}[tbp]%
  \begin{minipage}[t]{0.475\textwidth}
	  \centering
		\includegraphics[width=\textwidth]{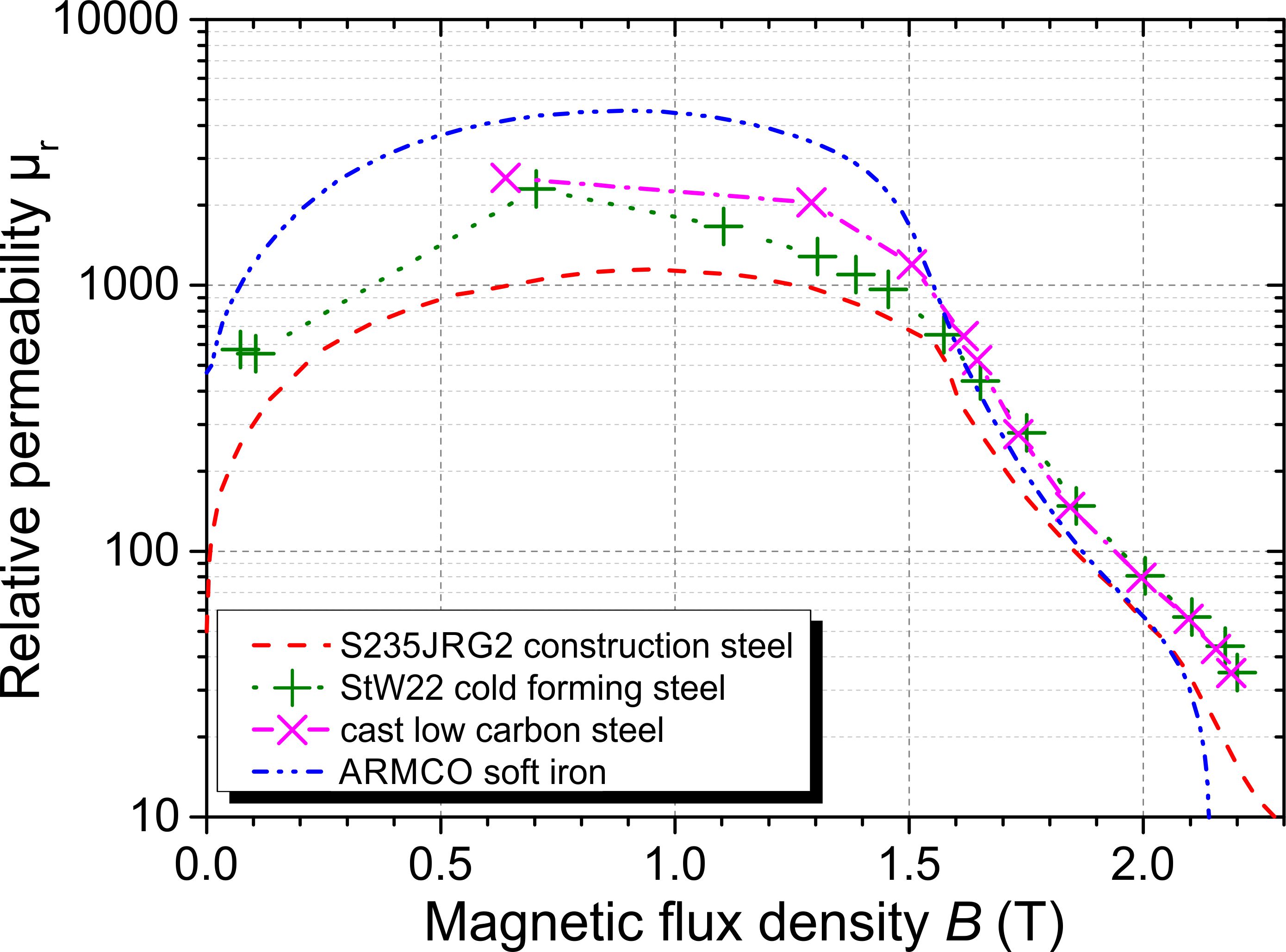}
		\caption{Permeability ($\mu_{\rm r}-B$) curves of four ferromagnetic materials:  S235JRG2 construction steel (dashed red) \cite{lederer:1998}, StW22 cold forming steel (dotted green) \cite{brammer:1995}, cast low carbon steel (dashed/dotted pink) \cite{ludwig:2006}, and ARMCO soft iron (dashed/dotted blue) \cite{spielvogel:2006}.}%
		\label{fig:MagnetizationCurve}%
	\end{minipage}
	\hfill
	\begin{minipage}[t]{0.475\textwidth}
		\centering
		\includegraphics[width=0.98\textwidth]{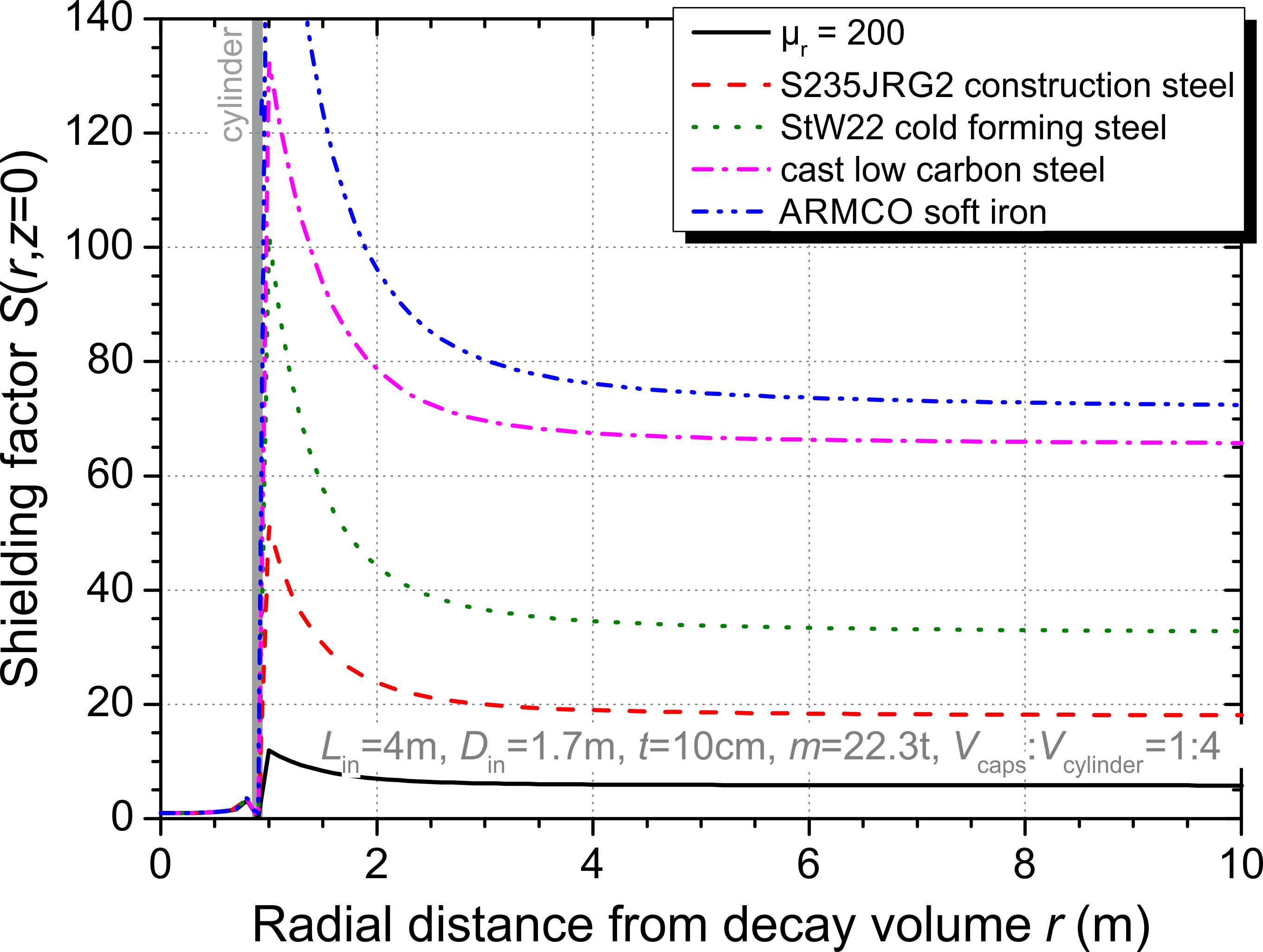}%
		\caption{Shielding factor $S(\rho,z=0)$ (cf. Eq.~(\ref{eq:S})) of a 10\,cm thick can (see the text and Fig.~\ref{fig:comsol2dstreamline} for details) surrounding the spectrometer \aspect, for the design current of $I_{\rm main}=100$\,A:  for constant relative permeability (black) and the materials shown in Fig.~\ref{fig:MagnetizationCurve}.  From a radial distance of 6\,m from the DV, the shielding factor becomes constant.}%
		\label{fig:ShieldingFactor}%
	\end{minipage}
\end{figure}

\begin{figure}[htb]%
	\begin{minipage}[t]{0.475\textwidth}
		\centering
		\includegraphics[width=0.87\textwidth]{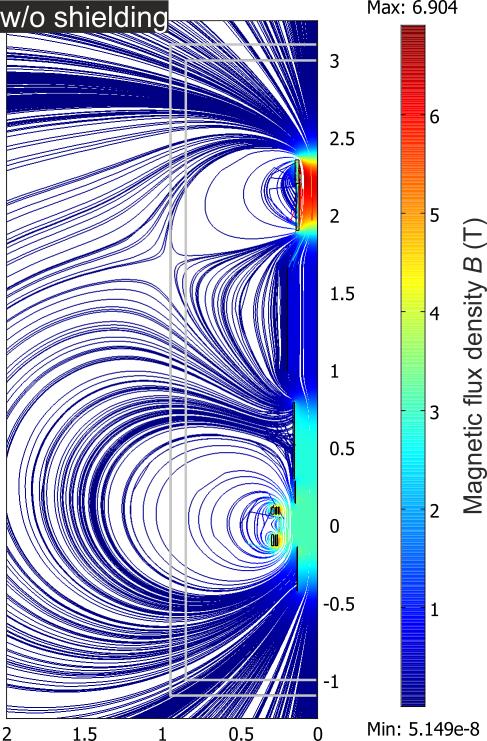}%
	\end{minipage}
	\hfill
	\begin{minipage}[t]{0.475\textwidth}
		\centering
		\includegraphics[width=0.87\textwidth]{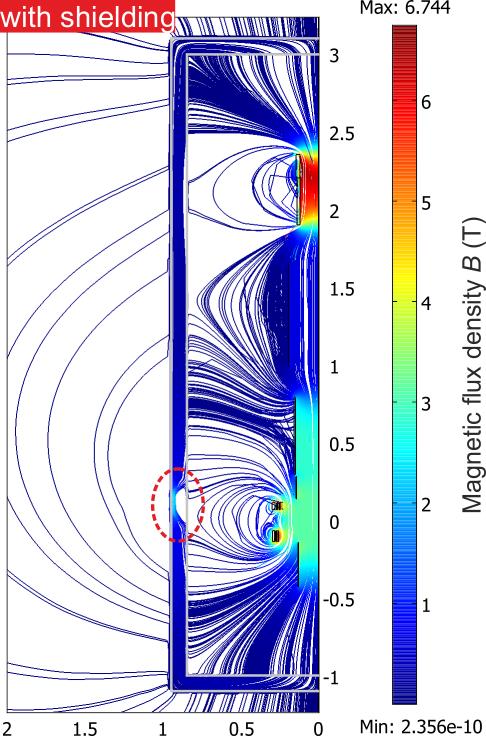}%
	\end{minipage}
	\caption{Streamline$^{\ref{foo:streamline}}$ plot of the magnetic flux density $\textbf{B}$ in the symmetry plane of a 2D axially symmetric shielding, for the {\aspect} design current of $I_{\rm main}=100$\,A.  Left:  Without magnetic shielding (boundaries drawn in gray), i.e., relative permeability $\mu_{\rm r}=1$, the coil system produces a strong stray magnetic field.   Right:  A can made of cast steel provides a reduction of the exterior magnetic field by a shielding factor of 66.  The dashed red ellipse highlights where the shielding cannot hold magnetic flux lines, i.e., where the cast iron is saturated (saturation flux density $B_{\rm s}\gtrsim1.28$\,T).}%
	\label{fig:comsol2dstreamline}%
\end{figure}

\begin{figure}[htbp]%
	\begin{minipage}[t]{0.475\textwidth}
		\centering
		\includegraphics[width=\textwidth]{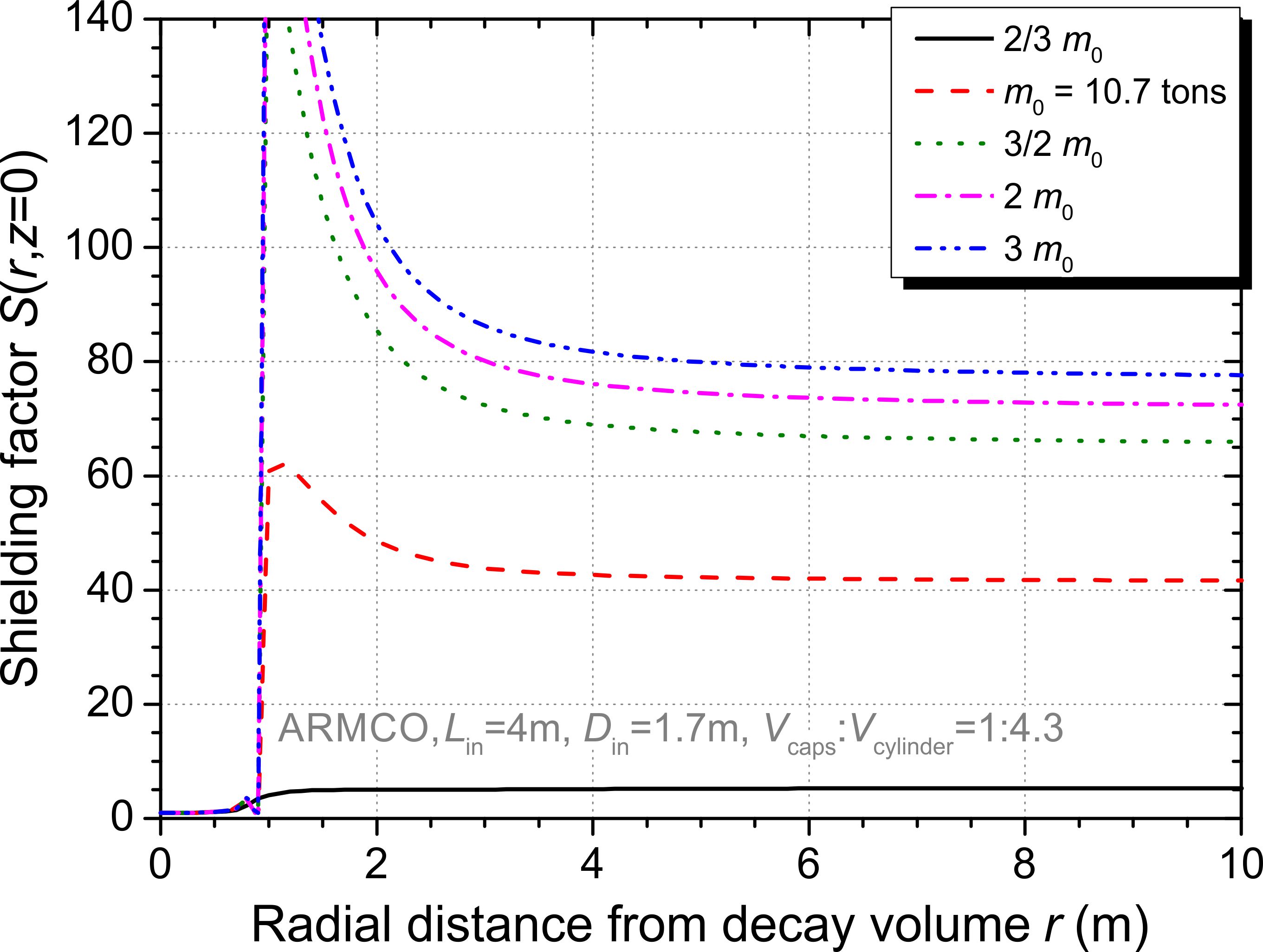}%
	\end{minipage}
	\hfill
	\begin{minipage}[t]{0.475\textwidth}
		\centering
		\includegraphics[width=\textwidth]{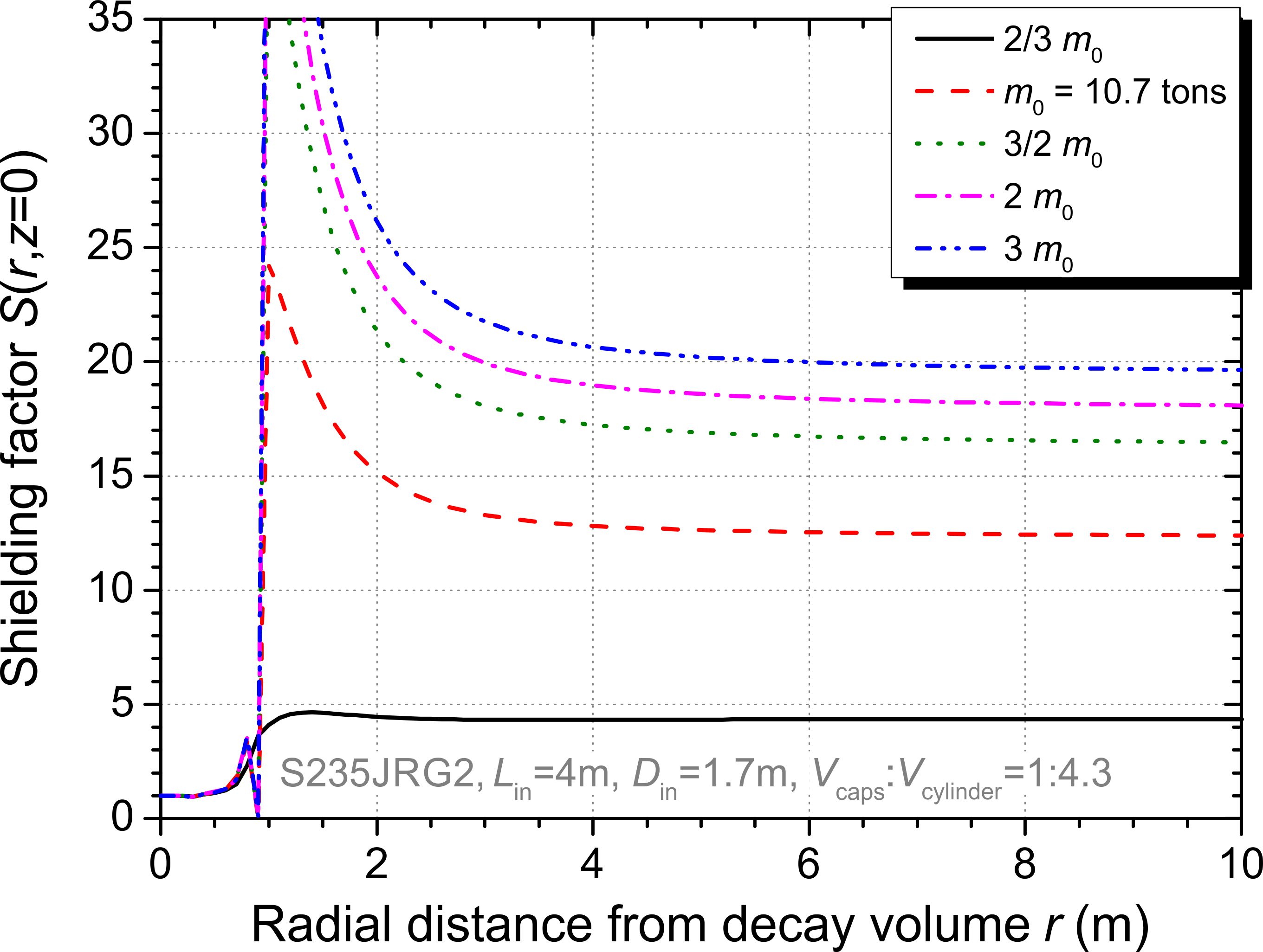}%
	\vspace{0.5cm}
	\end{minipage}
	\begin{minipage}[t]{0.475\textwidth}
		\centering
		\includegraphics[width=0.99\textwidth]{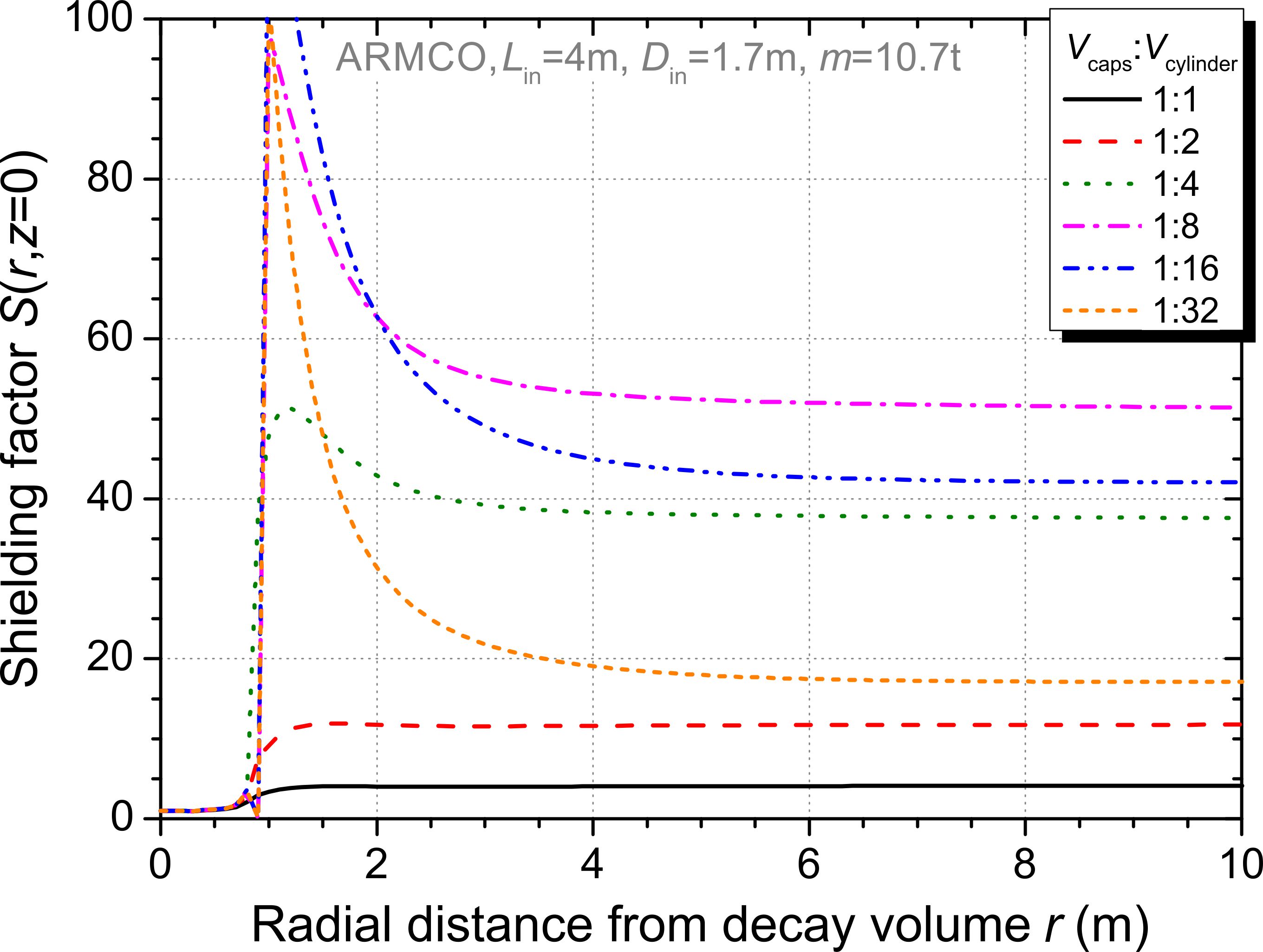}%
	\end{minipage}
	\hfill
	\begin{minipage}[t]{0.475\textwidth}
		\centering
		\includegraphics[width=\textwidth]{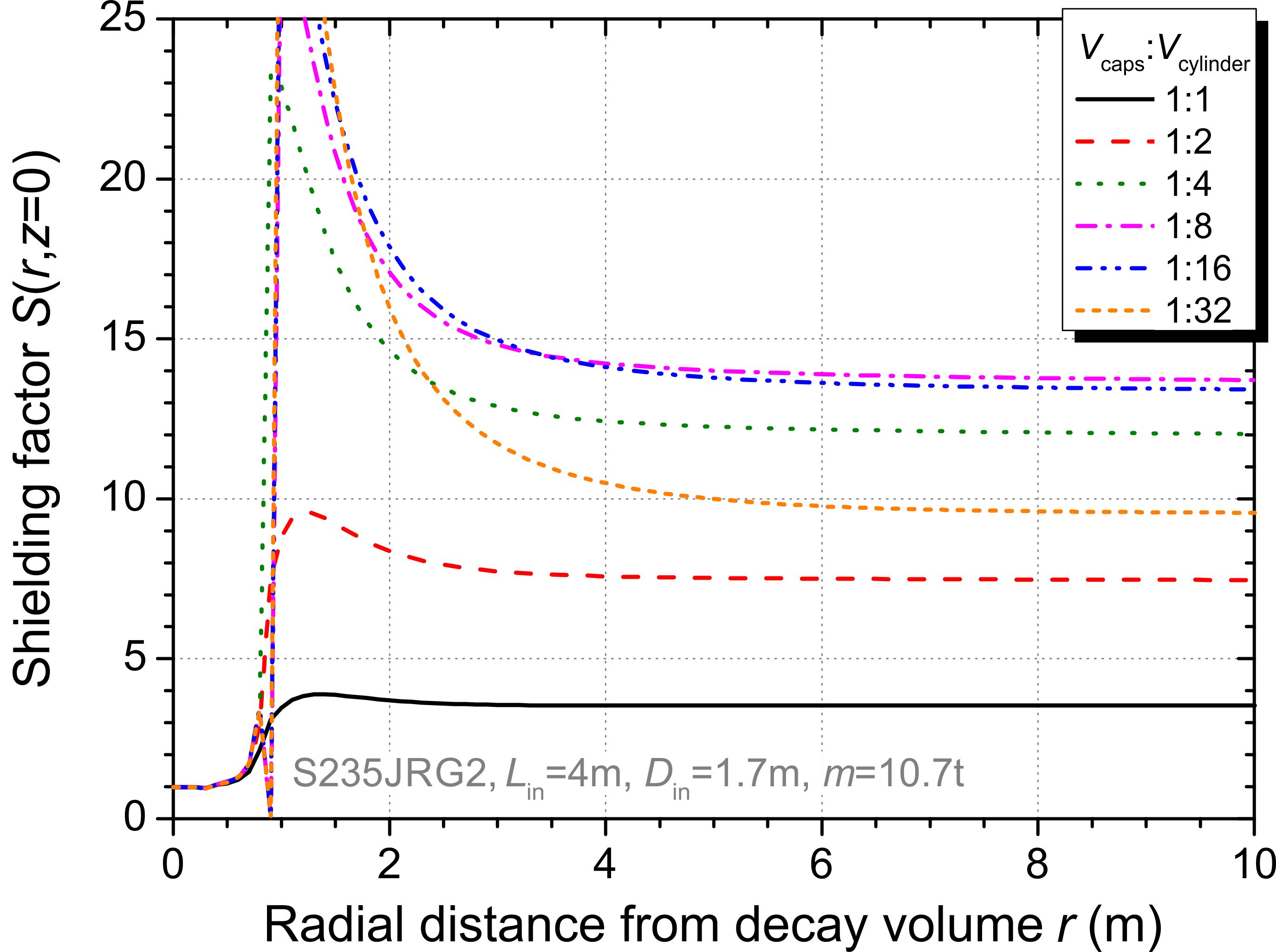}%
	\vspace{0.5cm}
	\end{minipage}
	\begin{minipage}[t]{0.475\textwidth}
		\centering
		\includegraphics[width=\textwidth]{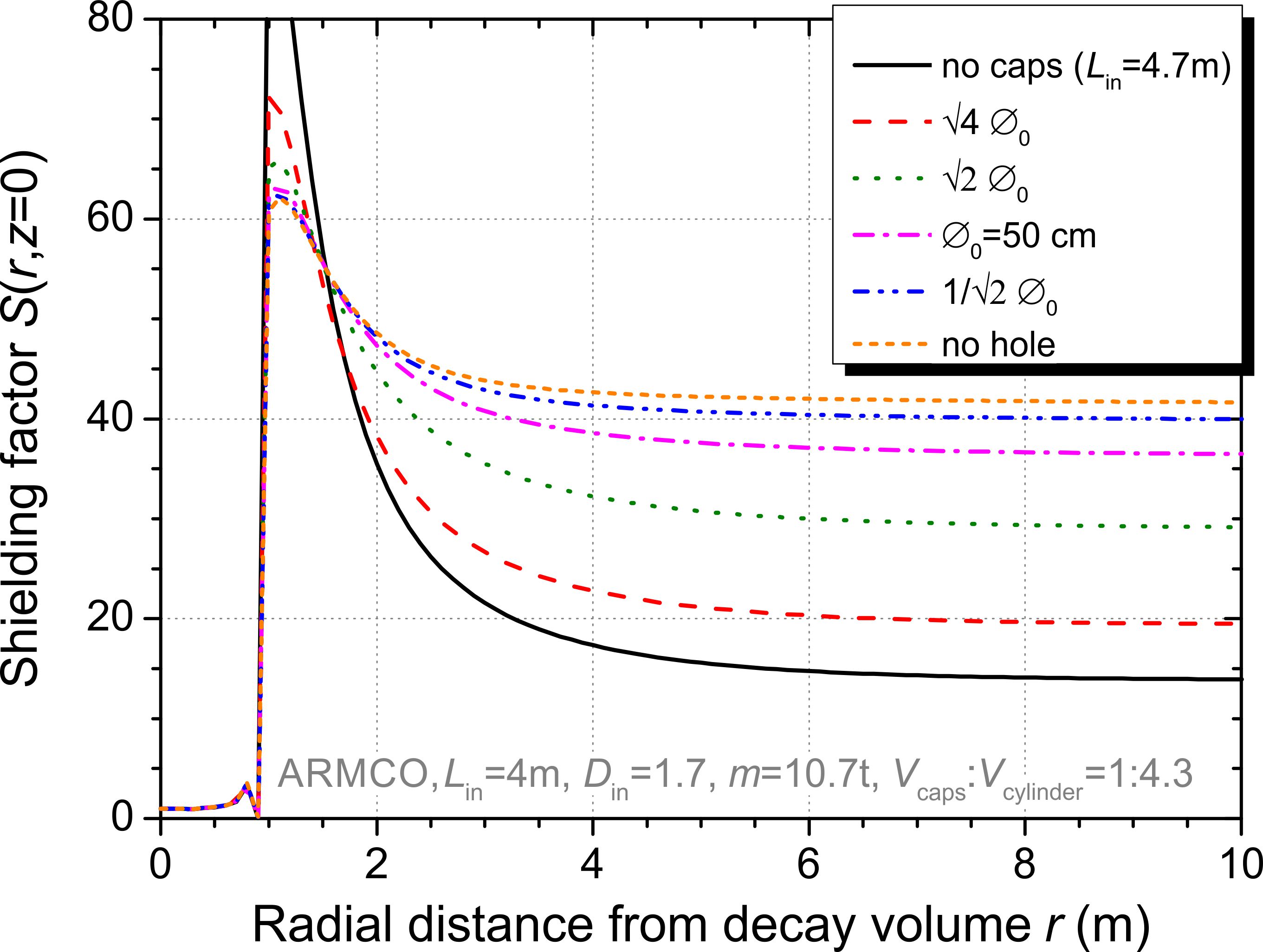}%
	\end{minipage}
	\hfill
	\begin{minipage}[t]{0.475\textwidth}
		\centering
		\includegraphics[width=\textwidth]{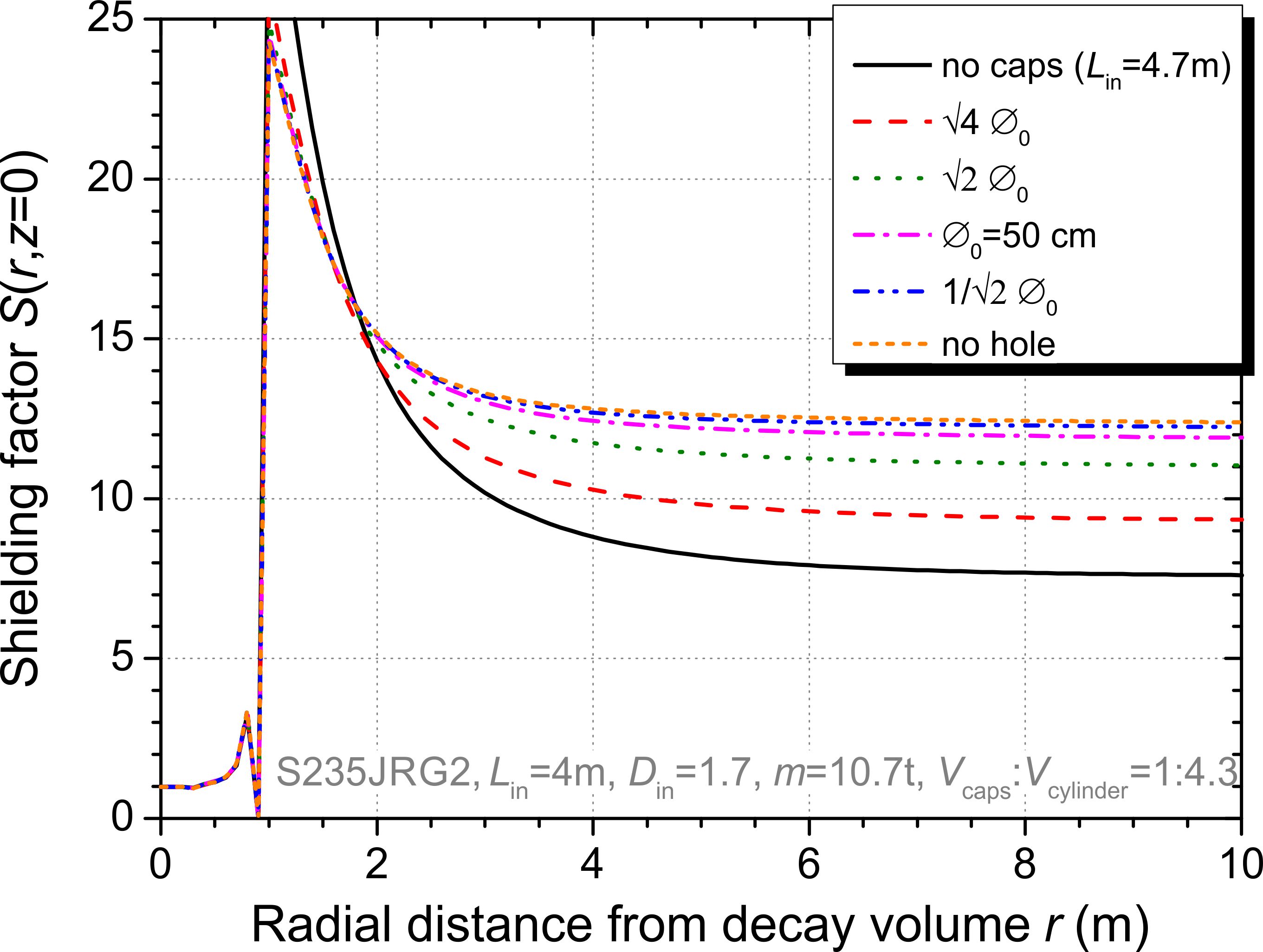}%
	\end{minipage}
	\caption{Influence of geometrical parameters on the shielding factor $S(\rho,z=0)$ (cf. Eq.~(\ref{eq:S})) for a can made of ARMCO iron (Left) or S235JRG2 steel (Right), respectively, surrounding the spectrometer \aspect, for the design current of $I_{\rm main}=100$\,A.  The six diagrams illustrate the correlation between shielding factor and shielding weight (Top), mass distribution of the shielding (Middle), or openings in the cap on top and on bottom (Bottom).  The top graphs show that a weight of about 10\,tons is sufficient to achieve a shielding factor of about 10.  For fixed weight (all except top), the strongest dependence exists on the volume ratio of top and bottom caps to cylinder (Middle).  Please note the different $y-$scales.}%
	\label{fig:ShieldingFactorSystematics}%
\end{figure}

Primarily, the simulations in 2D have served to verify and validate the FEM simulations by comparing their results with those of our numerical calculations.  Secondly, the 2D simulations have been used to investigate the correlation between shielding factor (defined by Eq.~(\ref{eq:S})) on the one hand and shielding material, weight, proportion, mass distribution, or openings on the other, cf. Figs.~\ref{fig:ShieldingFactor} and \ref{fig:ShieldingFactorSystematics}.  Finally, the simulations in 2D have served to gain a better understanding of magnetic saturation, cf. Fig.~\ref{fig:comsol2dstreamline}.

The 2D simulations have been performed in the COMSOL \textit{Azimuthal Induction Currents, Vector Potential} application mode.  In this application mode the PDE
\begin{eqnarray}
  \nabla \times \left( \frac{1}{\mu_0 \mu_{\rm r}} \nabla \times \textbf{A} - \textbf{M} \right) & = & J^{\rm e}_\varphi \hat{\textbf{e}}_\varphi 
\label{eq:PDE_Aphi}
\end{eqnarray}
with the \textit{Magnetic insulation} boundary condition
\begin{eqnarray}
  A_\varphi &=& 0
\label{eq:2D_BoundaryCondition}
\end{eqnarray}
is solved for the magnetic vector potential $\textbf{A}= A_\varphi \hat{\textbf{e}}_\varphi$, where $\hat{\textbf{e}}_\varphi$ is the (cylindrical) azimuthal angle unit vector.  In 2D, the Coulomb gauge fixing condition
\begin{eqnarray}
  \nabla \cdot \textbf{A} & = & 0.
\label{eq:GaugeFixing}
\end{eqnarray}
is always used. Here, $J^{\rm e}_\varphi$ is the current density, listed in Table~\ref{tab:coils}, generated externally by the coils c1 to c11.  Then the magnetization $\textbf{M}$ is zero and the vector potential $\textbf{A}$ is defined by
\begin{eqnarray}
  \textbf{B} = \mu_0 \mu_{\rm r} \textbf{H} & = & \nabla \times \textbf{A},
\label{eq:A}
\end{eqnarray}
where $\textbf{H}$ is the magnetic field strength.

On the other hand, we can replace the current density of a coil
\begin{eqnarray}
  \textbf{J} = J_\varphi \, \hat{\textbf{e}}_\varphi & = & \nabla \times \textbf{M}^{\rm e}
\label{eq:Je}
\end{eqnarray}
by an equivalent magnetization (see also Ref.~\cite{glueck:2011})
\begin{eqnarray}
  \textbf{M}^{\rm e} & = & J_\varphi \hat{\textbf{e}}_{\rm z} 
		\left\{ \begin{array}{l@{\quad \mbox{,} \quad}r}
		  		  	t  & \rho \leq r_{\rm in} \hphantom{,00} \\
						  t+r_{\rm in}-\rho & r_{\rm in} < \rho < r_{\rm in}+t \\
			 				0 & t+r_{\rm in} \leq \rho \hphantom{0000000}
		  			\end{array}
		\right.
\label{eq:M}
\end{eqnarray}
where $t$ is the radial thickness of the coil and $r_{\rm in}$ is its inner radius (cf. Table~\ref{tab:coils}).  Then we have to solve the PDE Eq.~(\ref{eq:PDE_Aphi}) with $\textbf{M}=\textbf{M}^{\rm e}$, $J^{\rm e}_\varphi=0$, and the boundary condition Eq.~\ref{eq:2D_BoundaryCondition} for the vector potential $\textbf{A}$, where Eq.~(\ref{eq:A}) is replaced by
\begin{eqnarray}
  \textbf{B} = \mu_0 \left( \textbf{H} + \textbf{M} \right) & = & \nabla \times \textbf{A}.
\label{eq:A_M}
\end{eqnarray}
For the purpose of comparison (cf. 3D application modes presented in Sec.~\ref{sec:nonaxially}), both the current density and the equivalent magnetization approach have been examined.

First we have verified that the results of the COMSOL simulations of the {\aspect} coil system\footnote{To achieve a fine mesh in COMSOL and at the same time high numerical accuracy, we simulated all but the tiny correction coils c3 and c5.} and of an axially symmetric shielding of constant\footnote{Please note that in Refs.~\cite{konrad:2007,konrad:2011c} the relative permeability is misstated as linear.} permeability are consistent with those of our numerical calculations.  For this, the coil system was surrounded with a can as described above.  The cylinder has a length of $L_{\rm in}=4$\,m, an inner diameter of $D_{\rm in}=1.7$\,m, and a thickness of $t=10$\,cm.  Both caps have a diameter of 1.9\,m and a thickness of 10\,cm, what corresponds to a total weight of $m=22.3$\,tons for iron/steel.

The permeability $\mu=\mu_0 \mu_{\rm r}=B H^{-1}$ (cf. Eq.~(\ref{eq:A})) of ferromagnetic materials is not constant but depends on $H$.  The dependence of the magnetic field $B$ on the magnetizing field $H$ is described by the magnetization curve, also called ${B-H}$ curve or hysteresis curve.  Different materials have different $B-H$ curves, as can be seen from Fig.~\ref{fig:MagnetizationCurve}.  Especially initial permeability, maximum permeability, and saturation strongly depend on the special kind of material (iron, construction steel, transformer steel, mu-metal, etc.), the thickness\footnote{In general, $B-H$ curves are measured for thin shielding material, i.e., for some tens of millimeters thick sheets.} of the material, and its pre- (and \mbox{post-)}treatment \cite{bozorth:1961,boll:1990}.  Hence we have completed the simulations in \textit{Axial symmetry} with 2D simulations of various shielding materials of non-linear permeability.  

The shielding factor $S(z)$ defined by
\begin{eqnarray}
  S(z) = \lim\limits_{\rho \rightarrow \infty}{S(\rho,z)} & = & \lim\limits_{\rho \rightarrow \infty}{\frac{B_{\mbox{w/o shield}}(\rho,z)}{B_{\mbox{with shield}}(\rho,z)}}
\label{eq:S}
\end{eqnarray}
is used as a measure for the effectiveness of the magnetic shielding.  Figure~\ref{fig:ShieldingFactor} shows the shielding factor $S(\rho,z=0)$ of a can as described above made of various shielding materials.  The shielding factor strongly depends on the special kind of material.  But from a radial distance of $\rho=6$\,m from the DV, the shielding factor becomes constant, as expected for physics reasons:  Shielded or unshielded, the dipole approximation Eq.~(\ref{eq:Bshield}) is a good approximation to  the far field.  Therefore, the far field drops off as $1/r^3$, and the shielding factor $S(z)$ is nothing more than the ratio of the proportionality factors.  The maximum shielding factor of $S(z=0)=72$ is reached for ARMCO iron.

Figure~\ref{fig:comsol2dstreamline} demonstrates the effectiveness of a can as described above made of cast steel\footnote{The $B-H$ curve of cast steel is taken from Ref.~\cite{ludwig:2006}.  Hence, in Refs.~\cite{konrad:2007, konrad:2011c}, this material is referred to as the special steel grade RTM3.}.  The cast steel provides a return path for most of the stray magnetic flux lines\footnote{\label{foo:streamline}The starting points for the magnetic field lines (streamlines) are randomized, which makes them unevenly distributed, but emphasizes the effectiveness of the magnetic shielding.} of the {\aspect} coil system and so significantly reduces the exterior magnetic field, with the exception of the top and the bottom cap.  At the caps, the magnetic flux pops out the shielding.  In addition, Fig.~\ref{fig:comsol2dstreamline} (Right) illustrates that the magnetic flux lines at the height of the coils c4a to c4c (cf. Figs.~\ref{fig:Sketch} and~\ref{fig:comsol3dmodel}) are forced to the outer radius of the cylinder.  The inner radius cannot hold magnetic flux lines since the cast iron is saturated.  This also explains why high-permeability alloys such as transformer steel or mu-metal are impractical for shielding the high stray magnetic field of \aspect. 

In order to study the influence of various geometrical parameters on the shielding factor, we have modeled a number of magnetic shieldings with the following geometry but variable parameters:   A can as described above of various inner length $L_{\rm in}$, inner diameter $D_{\rm in}$, cylinder thickness $t$, and cap thickness.  Both caps can have a centrical hole of variable diameter $\o$.

Figure~\ref{fig:ShieldingFactorSystematics} illustrates the correlation between shielding factor and shielding material, weight, mass distribution, or openings for a can made of ARMCO iron or S235JRG2 steel, respectively.  Firstly, it turned out that the strongest dependence exists on the volume ratio of top and bottom caps to cylinder (Middle).  The shielding factor is greatest for a mass distribution of 1:8.  But with increasing or decreasing mass distribution, the shielding factor decreases since the cylinder or caps are saturated, respectively.  Therefore, the mass distribution between caps and cylinder is fixed in the following.  The figure shows that a weight of about 10 tons with a mass distribution of 1:4 is sufficient to achieve a shielding factor of about 10 (Top), even with a hole in top and bottom cap with a diameter of 50\,cm (Bottom).  For decreasing length, the magnetic shielding becomes more compact and thus the shielding factor grows, almost independent of the diameter.  However, a diameter of $D_{\rm in}=1.7$\,m ideally matches a fixed length of $L_{\rm in}=4$\,m.

All in all, a closed geometry such as the can sufficiently reduces the stray magnetic field, but makes it very difficult to access the spectrometer {\aspect} in the magnetic field return yoke.  On the other hand, for a frame instead of the can, we expect a smaller but still sufficiently large enough shielding factor.  Thus, availability and price of shielding materials suggested to use a frame to reduce the exterior magnetic field of \aspect.  Such a non-axially symmetric geometry required FEM simulations in three dimensions, which are discussed in the following section~\ref{sec:nonaxially}.

As aforementioned in Sec.~\ref{sec:intro}, a return yoke made of ferromagnetic materials causes additional magnetic forces on the {\aspect} coils.  Requirements~\ref{i:R4} and~\ref{i:R5} therefore demand a minimum size for a passive shielding, and magnetic force calculations done parallel to the FEM simulations.  The force calculations are described in Sec.~\ref{sec:forces} below.

\section{3D non-axially symmetric shielding}
\label{sec:nonaxially}

We have continued to design a passive shielding for the spectrometer {\aspect} with COMSOL simulations in three dimensions.  Depending on the question of interest (shielding factor, field profile, field homogeneity, or magnetic forces), the simulations have been performed in the COMSOL \textit{Magnetostatics, Vector Potential} or in the \textit{Magnetostatics, No Currents} application mode.  In the \textit{Magnetostatics, Vector Potential} application mode the PDE
\begin{eqnarray}
  \nabla \times \left( \frac{1}{\mu_0 \mu_{\rm r}} \nabla \times \textbf{A} \right) & = & \textbf{J}^{\rm e}
\label{eq:PDE_A}
\end{eqnarray}
with the \textit{Magnetic insulation boundary condition
\begin{eqnarray}
  \textbf{n} \times \textbf{A} &=& 0
\label{eq:3DVector_BoundaryCondition}
\end{eqnarray}
}is solved for the magnetic vector potential $\textbf{A}$ defined by Eq.~(\ref{eq:A}) and for the gauge function $\psi$.  Here, 
\begin{eqnarray}
  \textbf{J}^{\rm e} = J^{\rm e}_\varphi \, \hat{\textbf{e}}_\varphi & = & J^{\rm e}_\varphi \left( \frac{y}{\sqrt{x^2+y^2}} \hat{\textbf{e}}_{\rm x} - \frac{x}{\sqrt{x^2+y^2}} \hat{\textbf{e}}_{\rm y} \right)
\label{eq:Jephi}
\end{eqnarray}
is the current density generated externally by the coils c1 to c11 (cf. Eq.~(\ref{eq:PDE_Aphi})), where $\hat{\textbf{e}}_{\rm x}$ and $\hat{\textbf{e}}_{\rm y}$ are the Cartesian unit vectors codirectional with the $x-$ and $y-$axes.  To avoid numerical instability, this application mode by default uses the Coulomb gauge fixing condition Eq.~(\ref{eq:GaugeFixing}).

In the COMSOL \textit{Magnetostatics, No Currents} application mode the PDE
\begin{eqnarray}
  -\nabla \cdot \left( \mu_0 \mu_{\rm r} \nabla \psi_{\rm m} - \mu_0 \textbf{M} \right) & = & 0
\label{eq:PDE_Vm}
\end{eqnarray}
with the \textit{Zero potential} boundary condition
\begin{eqnarray}
  \psi_{\rm m} &=& 0
\label{eq:3DScalar_BoundaryCondition}
\end{eqnarray}
is solved for the magnetic scalar potential $\psi_{\rm m}$.  The scalar potential is defined by
\begin{eqnarray}
  \textbf{H} = \frac{1}{\mu_0} \textbf{B} - \textbf{M} & = & - \nabla \psi_{\rm m},
\label{eq:Vm}
\end{eqnarray}
where the magnetization of the coils \textbf{M} is given by Eq.~(\ref{eq:M}).

\begin{figure}[t]%
	\begin{minipage}[t]{0.475\textwidth}
		\centering
		\includegraphics[width=0.72\textwidth]{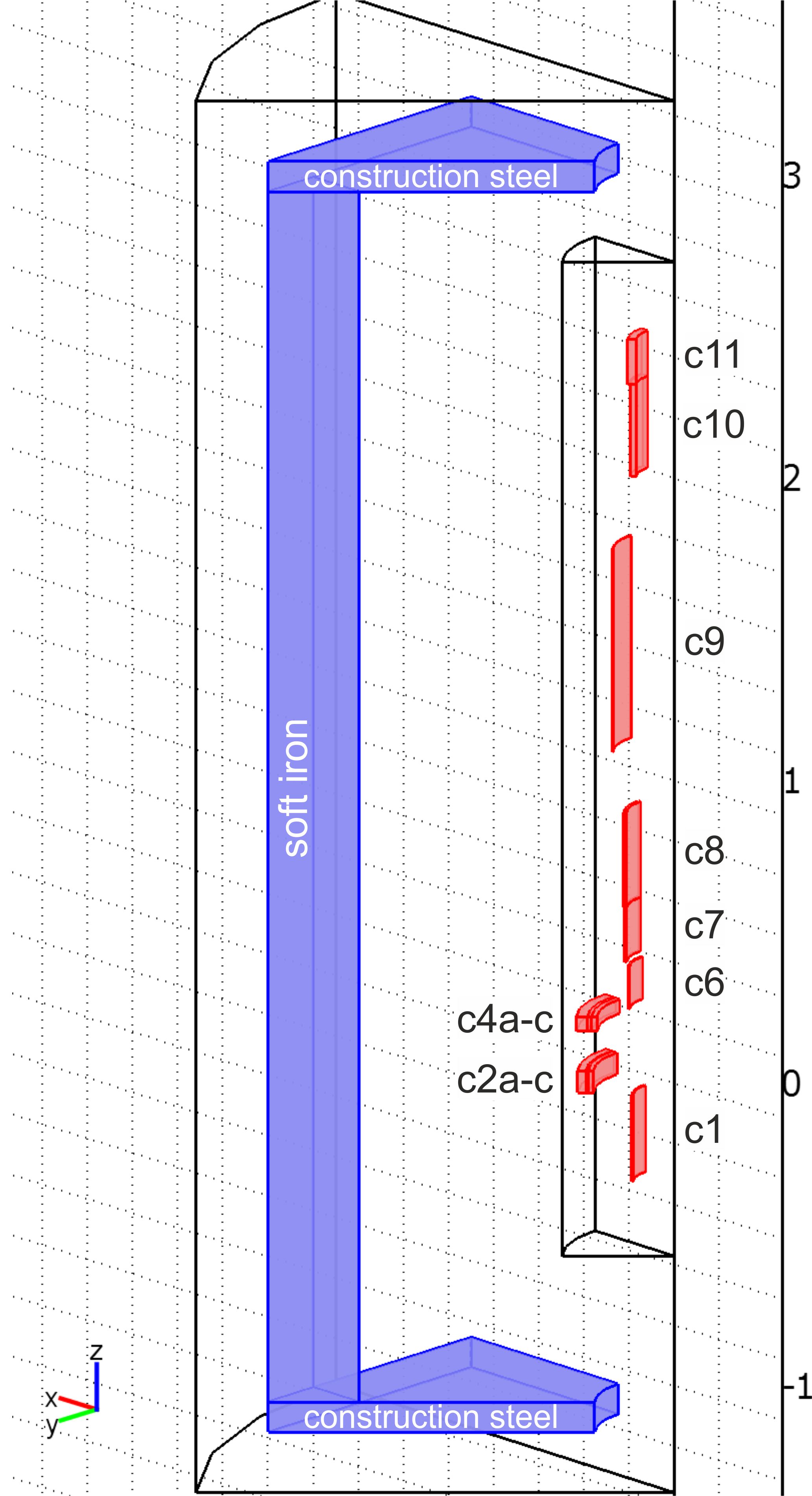}%
		\caption{Simulation model of the final design (cf. Fig.~\ref{fig:photo}) of a passive shielding for the spectrometer \aspect:  A combination of soft iron (4 billets) and construction steel (bottom and top plate).  To increase the numerical accuracy, we simulated only an eighth of the geometry by taking advantage of the symmetry of the magnet coils, denoted as c1 to c11, and of the shielding.}%
		\label{fig:comsol3dmodel}%
  \end{minipage}
	\hfill
	\begin{minipage}[t]{0.475\textwidth}
		\centering
		\includegraphics[width=0.832\textwidth]{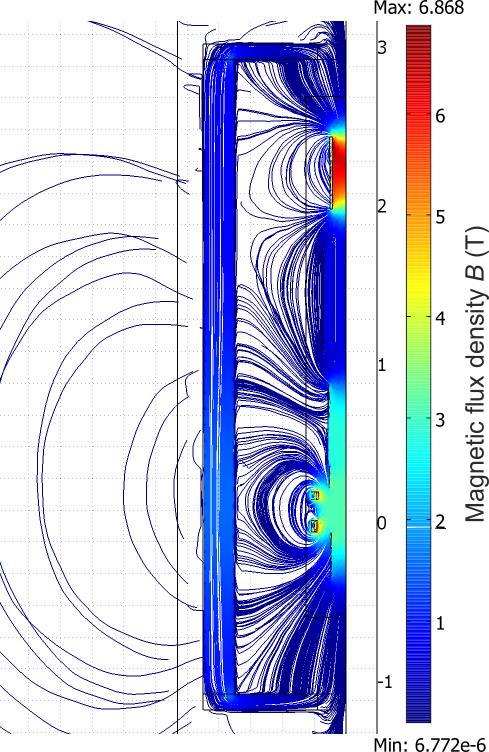}%
		\caption{Streamline$^{\ref{foo:streamline}}$ plot of the magnetic flux density $\textbf{B}$ in the symmetry plane through a billet of the simulation model shown in Fig.~\ref{fig:comsol3dmodel}, for the {\aspect} design current of $I_{\rm main}=100$\,A.  The magnetic shielding ensures a sufficient reduction of the exterior magnetic field without significant influence on shape and homogeneity of the internal magnetic field, as can be seen from Figs.~\ref{fig:DVshielding} and~\ref{fig:APshielding}.}%
		\label{fig:comsol3dstreamline}%
  \end{minipage}
\end{figure}

In order to assess how well the requirements \ref{i:R1}. and \ref{i:R3}. to \ref{i:R5}. (cf. Sec.~\ref{sec:intro}) are satisfied, a fine mesh in radial direction (up to at least 10\,m from the DV), close to the symmetry axis of the magnet, in the DV, in the AP, and within the {\aspect} coils is demanded, and all at the same time.  This condition is equivalent to a fine mesh in the entire space and therefore almost impossible with the software used.  Nevertheless, in order to refine the mesh in our regions of interest (radial direction, symmetry axis of the magnet, DV, AP, and coils), we simulated only an eighth (45$^\circ$ circular segment) of the geometry by taking advantage of the symmetry of the coil system and of the magnetic shielding.  Figure~\ref{fig:comsol3dmodel} shows that the radial direction transforms into two boundaries (crop margins), the symmetry axis of the magnet into an edge ($z-$axis) of the geometry, and the volume of the coils decreases considerably.  Therefore one has to refine the mesh in both symmetry planes (crop margins), along the $z-$axis, and within the coils only.  In this way, the computing time is reduced and at the same time the numerical accuracy increased.

First we have checked that also the results of 3D simulations of axially symmetric shieldings are consistent with those of the corresponding 2D simulations (for details see Sec.~\ref{sec:axially}).  Then we have completed the design of a passive shielding for the spectrometer {\aspect} with 3D simulations of various geometries and combinations of shielding materials of non-linear permeability (cf. Fig.~\ref{fig:ShieldingFactor} and Sec.~\ref{sec:axially}).  We modeled a number of magnetic shieldings with the following geometry but variable parameters:  A frame consisting of a bottom and a top plate and $N \times 4$ billets between the two plates, $N$ billets in each of the four corners of the plates.  Both plates are of variable cross-section and thickness, and have a centrical hole with variable diameter $\o$.  Each quartet of billets is of variable cross-section and length. 

On both symmetry planes (crop margins), the magnetic field is tangential to the plane.  This is described by the \textit{Magnetic insulation} boundary condition
\begin{eqnarray}
  \textbf{n} \times \textbf{A} & = & 0 \quad , \quad \mbox{for the \textit{Magnetostatics, Vector Potential} application mode} \nonumber \\
  \textbf{n} \cdot \textbf{B} & = & 0 \quad , \quad \mbox{for the \textit{Magnetostatics, No Currents} application mode} \nonumber
\label{eq:BoundaryCondition}
\end{eqnarray} 

Figure~\ref{fig:comsol3dmodel} shows the final design of the magnetic field return yoke:  The {\aspect} coil system is surrounded with a frame made of soft iron (4 billets) and construction steel (bottom and top plate).  Each plate has a volume of $18 \times 18 \times 1$\,dm$^3=324$\,liters, minus a hole with a diameter of 50\,cm (20\,liters), and each billet has a volume of $2 \times 2 \times 40$\,dm$^3=160$\,liters, what corresponds to a total weight of 9.8\,tons for iron/steel.  We have chosen a combination of ARMCO iron (billets) and S235JRG2\footnote{The $B-H$ curve of S235JRG2 (former name: St37-2) construction steel is taken from Ref.~\cite{lederer:1998}.} steel (plates), only because of short time of delivery and price of shielding materials.

Figure~\ref{fig:comsol3dstreamline} demonstrates the effectiveness of the final design of the return yoke shown in Fig.~\ref{fig:comsol3dmodel}.  Compared to Fig.~\ref{fig:comsol2dstreamline} (Right), the magnetic flux lines$^{\ref{foo:streamline}}$ do not exit the plates as the caps before.  Both holes in the bottom and in the top plate ensure that the return yoke provides a complete path for the stray magnetic flux lines.  And, in contrast to the cylinder, the ARMCO iron billets are not saturated (saturation flux density $B_{\rm s}=2.13$\,T):  The magnetic flux lines are spread across the entire thickness of the billets.

All in all, the final design meets the requirements~\ref{i:R1} to~\ref{i:R3} (cf. Sec.~\ref{sec:intro}):  The stray magnetic field is suppressed to less than 1\,Gauss (0.1\,mT) in a radial distance of 5\,m from the DV, as can be seen from Fig.~\ref{fig:radialdata} below.  The cardiac pacemaker limit of 5\,Gauss (0.5\,mT) is reached at a radial distance of 2.8\,m from the DV, for the design current of $I_{\rm main}=100$\,A, i.e., at 2.5\,m for our working current of $I_{\rm main}=70$\,A.  From a radial distance of $\rho=10$\,m from the DV, the shielding factor $S(\rho,\varphi,z)$ becomes constant and equal to $S(\varphi=0^\circ,z=0)=8.3$.  This has to be compared with a shielding factor of 12.4 (41.5) for a 5\,cm thick can as described in Sec.~\ref{sec:axially} made of S235JRG2 steel (ARMCO iron), cf. Fig.~\ref{fig:ShieldingFactorSystematics}.  It should be noted that, for steel billets, ARMCO iron instead of steel plates marginally increase the shielding factor to 12.6, whereas, for ARMCO billets, steel instead of ARMCO iron plates slightly reduce the shielding factor to 38.8.  But it is evident that an open instead of a closed geometry with a weight of 9.8 instead of 10.7\,tons and a mass distribution of 1:1 instead of 1:4.3 comes with a significant but tolerable reduction in shielding factor (see also Fig.~\ref{fig:ShieldingFactorSystematics}).  Comparing to Fig.~\ref{fig:ShieldingFactorSystematics} (Middle), the shielding factor $S(\rho,\varphi=0^\circ,z=0)$ behaves like for a mass distribution of 1:1, i.e., it is monotonically increasing, but does not drop below 5.

\begin{figure}[t]%
	\begin{minipage}[t]{0.475\textwidth}
  	\includegraphics[width=\textwidth]{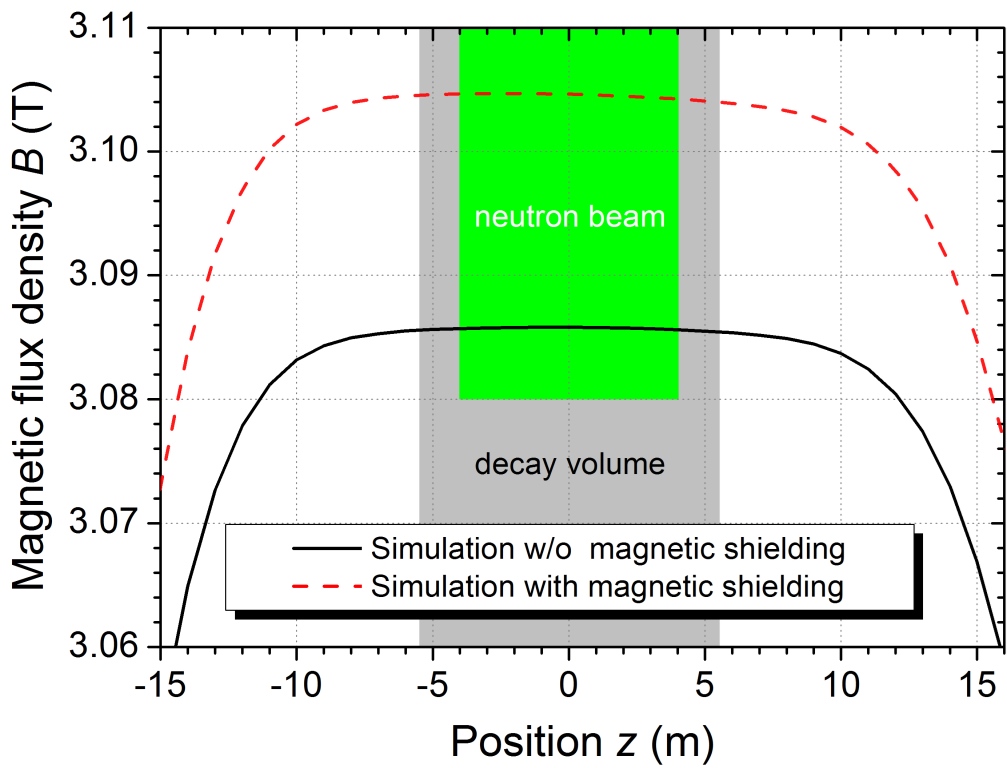}%
		\caption{Calculated influence of the magnetic shielding on the field in the decay volume (gray bar) of the spectrometer \aspect:  Magnetic flux density $B(z)$ along the $z-$axis of the simulation model shown in Fig.~\ref{fig:comsol3dmodel}, without (black) and with shielding (dashed red), for the design current of $I_{\rm main}=100$\,A.  Shape and high homogeneity of the internal magnetic field remain unaffected, while its height changes by 0.6\,\%.}%
		\label{fig:DVshielding}%
  \end{minipage}
	\hfill
	\begin{minipage}[t]{0.475\textwidth}
	  \includegraphics[width=\textwidth]{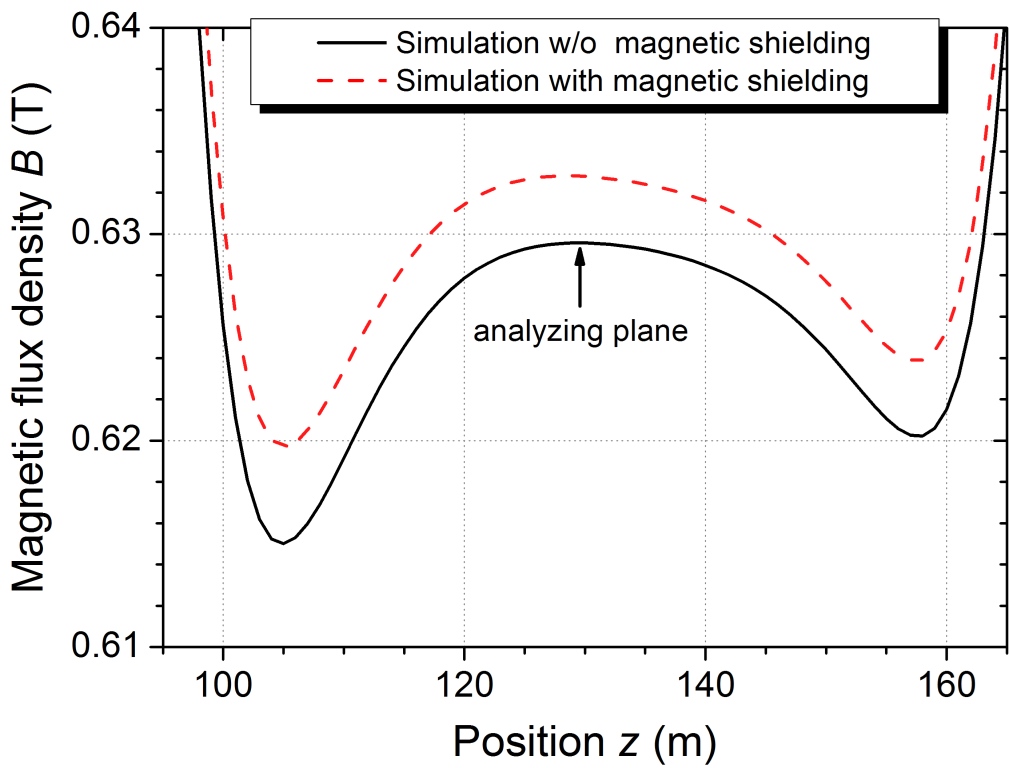}%
		\caption{Calculated influence of the magnetic shielding on the field in the analyzing plane:  Magnetic flux density $B(z)$ along the $z-$axis of the simulation model shown in Fig.~\ref{fig:comsol3dmodel}, without (black) and with shielding (dashed red), for the {\aspect} design current of $I_{\rm main}=100$\,A.  Shape and high homogeneity of the internal magnetic field remain unaffected, while its height changes by 0.5\,\%.}%
		\label{fig:APshielding}%
  \end{minipage}
\end{figure}

In addition, the influence of the return yoke on the internal magnetic field is quite small, as can be seen from Figs.~\ref{fig:DVshielding} and~\ref{fig:APshielding}.\footnote{\label{foo:SimulatedData}To increase the numerical accuracy of the calculations, the magnetic field has been simulated for real permeability and for relative permeability set to $\mu_{\rm r}=1$, and their difference has been added to our 2D axially symmetric, numerical calculations for $\mu_{\rm r}=1$.}  Both the shape and the high homogeneity of the magnetic field in the DV and in the AP remain unaffected, while the magnetic field ratio $r_{\rm B}$ (cf. Sec.~\ref{sec:aspect}) changes slightly by $-9\times10^{-4}$ (relative).  Such a ratio change needs to be taken into account in the extraction of the angular correlation coefficient $a$ from the measured proton count rates (for details see Ref.~\cite{glueck:2005}), but is otherwise irrelevant for the purpose of the {\aspect} experiment.

\section{Magnetic forces}
\label{sec:forces}

As aforementioned in Sec.~\ref{sec:axially}, the 3D simulations of a passive shielding were accompanied by magnetic force calculations:  The magnetic force $\textbf{F}_i$ on coil c$i$ ($i=1, 2a,\ldots, 11$) caused by an external magnetic field $\textbf{B}_{\rm ext}$ is given by
\begin{eqnarray}
  \textbf{F}_{i} & = & \int\limits_{V_i} {\rm d}^3r \, \textbf{J}_i(\textbf{r}) \times \textbf{B}_{\rm ext}(\textbf{r}) ,
\label{eq:Fiext}
\end{eqnarray}
where $V_i$ is the coil volume and $\textbf{J}_i$ its current density (see also Table~\ref{tab:coils}).  The magnetic force calculations have therefore been performed in the COMSOL \textit{Magnetostatics, Vector Potential} application mode\footnote{Please note that in Refs.~\cite{konrad:2007,konrad:2011c} the application mode is misstated.} (for details see Sec.~\ref{sec:nonaxially}), through \textit{Subdomain Integration} in the COMSOL \textit{Postprocessing} mode.  In our case, the external magnetic field $\textbf{B}_{\rm ext}$ is composed of the magnetic fields $\textbf{B}_j$ of the other coils c$j$ ($j\neq i$) and the magnetic field $\textbf{B}_{\rm shield}$ caused by the magnetic field return yoke:
\begin{eqnarray}
   \textbf{B}_{\rm ext}(\textbf{r}) & = & \sum_{j\neq i} \textbf{B}_j(\textbf{r}) + \textbf{B}_{\rm shield}(\textbf{r}) .
\label{eq:Bext}
\end{eqnarray}
Hence, the force $\textbf{F}_i$ is composed of the forces $\textbf{F}_{ij}$ caused by the other coils and the force $\textbf{F}_{i,\rm shield}$ caused by the return yoke:
\begin{eqnarray}
   \textbf{F}_i & = & \sum_{j\neq i} \underbrace{\int\limits_{V_i} {\rm d}^3r \, \textbf{J}_i(\textbf{r}) \times \textbf{B}_j(\textbf{r})}_{\textbf{F}_{ij}} + \underbrace{\int\limits_{V_i} {\rm d}^3r \, \textbf{J}_i(\textbf{r}) \times \textbf{B}_{\rm shield}(\textbf{r})}_{\textbf{F}_{i,\rm shield}}.
\label{eq:Fi}
\end{eqnarray}
The additional magnetic forces $\textbf{F}_{i,\rm shield}$ on the {\aspect} coils are thus calculated as the force differences between a simulation with and one without magnetic shielding.  To increase the numerical accuracy of the calculations, the latter is replaced by the simulation with magnetic shielding in which the relative permeability of the shielding materials is set to $\mu_{\rm r}=1$.

Table~\ref{tab:forces} lists the magnetic forces on an eighth of the {\aspect} coils for the simulation model shown in Fig.~\ref{fig:comsol3dmodel}.  To simplify interpretation, the two Cartesian coordinates $F_{\rm x}$ and $F_{\rm y}$ are converted to the polar coordinates $F_\rho = \sqrt{F_{\rm x}^2+F_{\rm y}^2}$ and $\varphi = \arcsin \frac{F_{\rm y}}{F_\rho}$.  As was required (cf. condition~\ref{i:R4} in Sec.~\ref{sec:intro}), none of the forces changes its sign due to the return yoke.  In addition, the relative force changes are quite small, with the exception of the already small radial forces $F_\rho$ on the coils c2c and c4c.

\begin{table}[t]%
\caption{Magnetic forces on an eighth of the {\aspect} coils for the simulation model shown in Fig.~\ref{fig:comsol3dmodel}.  The first column specifies the description of the coil.  The second and third column list the radial and the axial forces caused by the other coils, in units of the supply current in amperes.  The fourth column gives the angle $\varphi$ of the radial forces.  The fifth and sixth column indicate the radial and the axial forces caused by the magnetic field return yoke, at the design current of $I_{\rm main}=100$\,A.  Data taken from Ref.~\cite{konrad:2007}.  The last column gives the angle changes $\delta\varphi$ caused by the return yoke.}
\centering
\begin{tabular}{l|rr|r|rr|r}
  \hline
  coil & $F_\rho I_{\rm main}^{-2}$ (N\,A$^{-2}$) & $F_{\rm z} I_{\rm main}^{-2}$ (N\,A$^{-2}$) & $\varphi$ ($^\circ$) & $F_{\rho\rm,shield} I_{\rm main}^{-2}$ (N\,A$^{-2}$) & $F_{\rm z,shield} I_{\rm main}^{-2}$ (N\,A$^{-2}$) & $\delta\varphi$ ($^\circ$) \\
  \hline
  c1 & 6.5\hphantom{0} & 1.5\hphantom{00} & 22.5 & 0.05 & -0.002 & 0.0 \\
  c2a & 14.3\hphantom{0} & 1.7\hphantom{00} & 22.5 & 0.05 & -0.003 & 0.0 \\
  c2b & 6.1\hphantom{0} & 1.0\hphantom{00} & 22.5 & 0.03 & -0.002 & 0.0 \\
  c2c & 0.6\hphantom{0} & 5.3\hphantom{00} & 22.5 & -0.14 & -0.010 & 0.0 \\
  c4a & 10.0\hphantom{0} & -1.0\hphantom{00} & 22.5 & 0.03 & -0.002 & 0.0 \\
  c4b & 7.3\hphantom{0} & -1.1\hphantom{00} & 22.5 & 0.03 & -0.002 & 0.0 \\
  c4c & 0.03 & -4.4\hphantom{00} & 16.8 & 0.12 & -0.009 & 4.4 \\
  c6 & 4.4\hphantom{0} & -0.4\hphantom{00} & 22.5 & 0.02 & -0.001 & 0.0 \\
  c7 & 6.6\hphantom{0} & 0.2\hphantom{00} & 22.5 & 0.04 & -0.002 & 0.0 \\
  c8 & 10.9\hphantom{0} & -2.6\hphantom{00} & 22.5 & 0.07 & -0.005 & 0.0 \\
  c9 & 1.6\hphantom{0} & 0.003 & 22.5 & 0.03 & 0\hphantom{.000} & 0.0 \\
  c10 & 37.9\hphantom{0} & 10.4\hphantom{00} & 22.4 & 0.16 & 0.020 & 0.0 \\
  c11 & 25.6\hphantom{0} & -10.5\hphantom{00} & 22.5 & 0.16 & 0.020 & 0.0 \\
  \hline
  sum & & 0.1\hphantom{00} & & & 0\hphantom{.000} \\
  \hline  
\end{tabular}
\label{tab:forces}
\end{table}

Without magnetic shielding, the sum of the axial forces $F_{\rm z}$ on the {\aspect} coils must be zero, in contrast to the calculated sum of 1\,kN at the design current of $I_{\rm main}=100$\,A (cf. Table~\ref{tab:forces}).  Moreover, the (azimuthal) angle $\varphi$ of the radial forces on an eighth of the coils (45$^\circ$ circular segments) must be 22.5$^\circ$, in accordance with the calculated angles, with the exception of coil c4c in which small numerical errors in the small components $F_{\rm x}$ and $F_{\rm y}$ explain a deviation in $\varphi$ of 5.7$^\circ$ respectively 1.3$^\circ$.  However, numerical errors of this (relative) magnitude can be tolerated.

\begin{figure}[tbp]%
\centering
\includegraphics[width=0.475\textwidth]{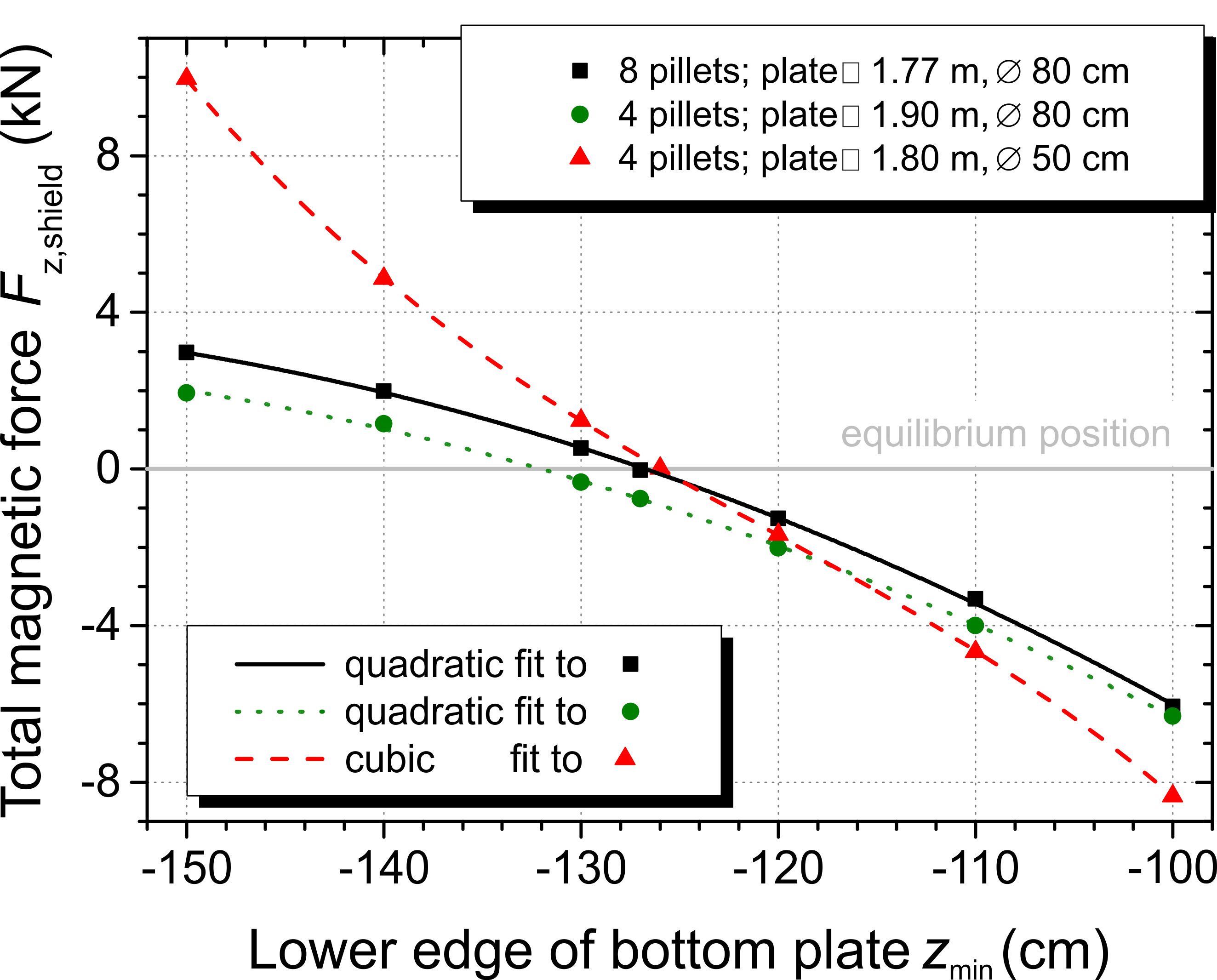}%
\caption{Sum of the additional axial forces $F_{\rm z,shield}$ on the {\aspect} coil system, for the design current of $I_{\rm main}=100$\,A.  To determine the equilibrium position (gray line), the return yoke is shifted along the symmetry axis of the magnet.  For $z_{\rm min}=-113.5$\,cm, the coil system is geometrically but not magnetically centered within its magnetic shielding:  A frame made of $N\times4$ ARMCO iron billets (cross-section $20\times20$\,cm$^2$, length 4\,m) and two S235JRG2 steel plates (thickness 10\,cm) of variable cross-section $\square$, and with a centrical hole of variable diameter $\varnothing$.  To magnetically center, e.g., the final design shown in Fig.~\ref{fig:comsol3dmodel} (red triangles), the return yoke must be lowered by $-12.5$\,cm to $z_{\rm min}=-126$\,cm.}%
\label{fig:MagneticForces}%
\end{figure}

With magnetic shielding, the sum of the axial forces $F_{\rm z,shield}$ listed in Table~\ref{tab:forces} is zero, i.e., the {\aspect} coil system is already placed at the location in which the additional magnetic forces cancel out.  The equilibrium position is determined by shifting the magnetic shielding along the symmetry axis of the magnet, meanwhile the coil system is hold in position, until the sum of the forces is zero (cf. Fig.~\ref{fig:MagneticForces}).  In this way, also the axial force gradient due to the return yoke shown in Fig.~\ref{fig:comsol3dmodel} was fitted to
\begin{eqnarray}
  \frac{\delta F_{\rm z,shield}}{\delta z} & = & 291 \frac{\rm N}{\rm cm},
\label{eq:dFzdz}
\end{eqnarray}
at the design current of $I_{\rm main}=100$\,A.  According to Eq.~(\ref{eq:Fiext}), the axial force gradient scales almost with $I_{\rm main}^2$.

\section{Experimental results}
\label{sec:results}

\begin{figure}[htbp]%
	\centering
	\includegraphics[width=0.425\textwidth]{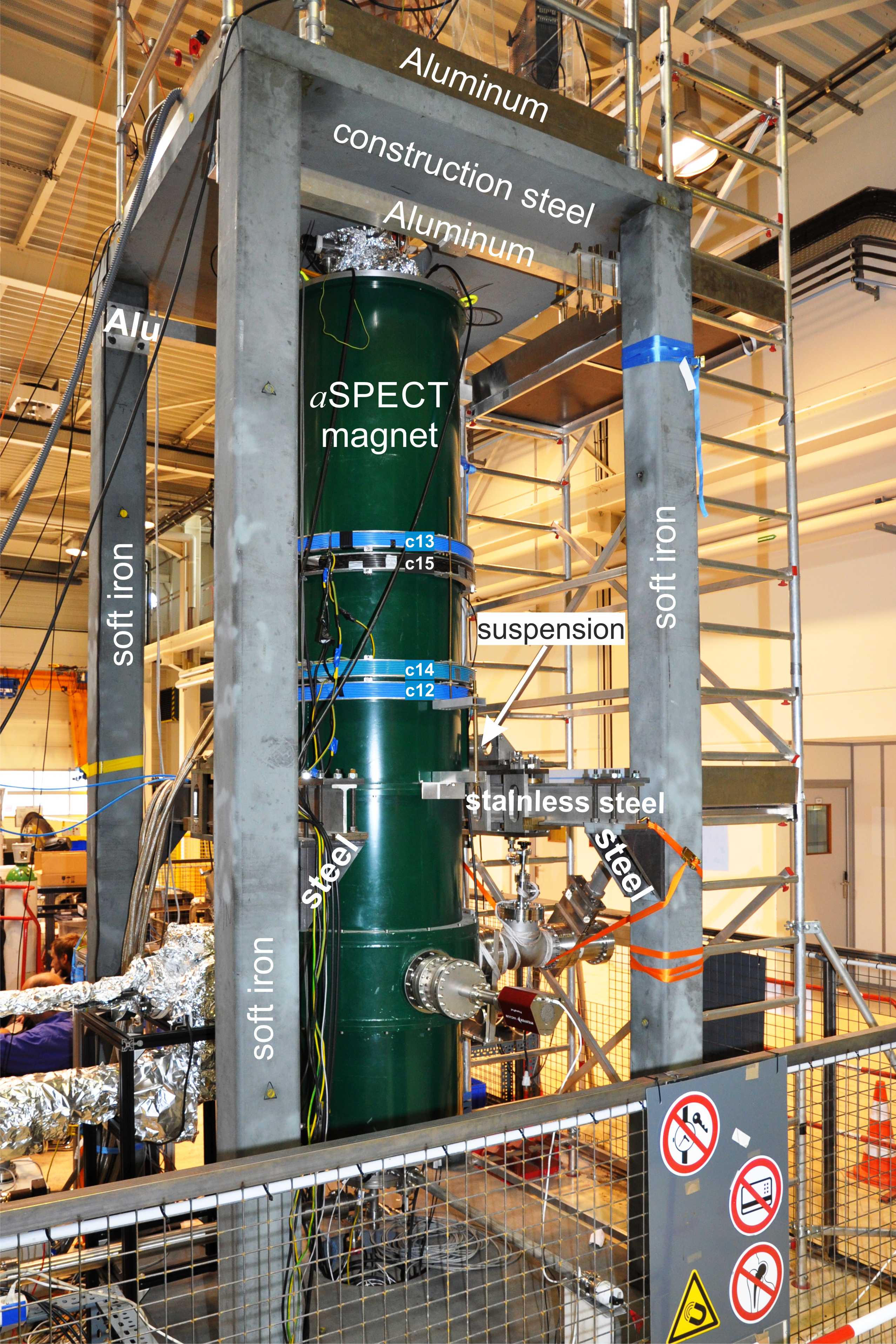}%
	\caption{Experimental set-up of the {\aspect} magnet within its return yoke at the ILL:  A structure made of soft iron (4 billets of cross-section $20\times20$\,cm$^2$ and length 4\,m) and construction steel (bottom and top plate of cross-section $1.8\times1.8$\,m$^2$ and thickness 10\,cm) serves to suppress the stray magnetic field to $<1$\,Gauss (cf. Fig.~\ref{fig:radialdata}) in a radial distance of 5\,m from the decay volume.  Photograph courtesy of A. Wunderle \cite{wunderle:2012}.}%
	\label{fig:photo}%
\end{figure}

First the spectrometer {\aspect} was centered within the magnetic field return yoke, shown in Fig.~\ref{fig:photo}, based on the magnetic force calculations (for details see Sec.~\ref{sec:forces}).  For this purpose, the magnet was hanging from a crane in its designed location and separated from its suspension (cf. Fig.~\ref{fig:photo}) to the return yoke, meanwhile the magnetic field was ramped up very slowly.  The unstable equilibrium position was determined with the aid of a spring balance, installed between the {\aspect} magnet and the crane hook, displaying the deviation from the magnet's own weight.  The position thus obtained deviates from the calculated equilibrium position by only 1\,cm along the symmetry axis of the magnet, which corresponds to a deviation of only 291\,N ($\approx 29$\,kg) from the magnetic force calculations (cf. Eq.~(\ref{eq:dFzdz})), or 143\,N ($\approx 14$\,kg) for the working current of $I_{\rm main}=70$\,A.  This result has been validated both at the TRIGA research reactor in Mainz, Germany and at the ILL.

\begin{figure}[htbp]%
	\centering
	\includegraphics[width=0.475\textwidth]{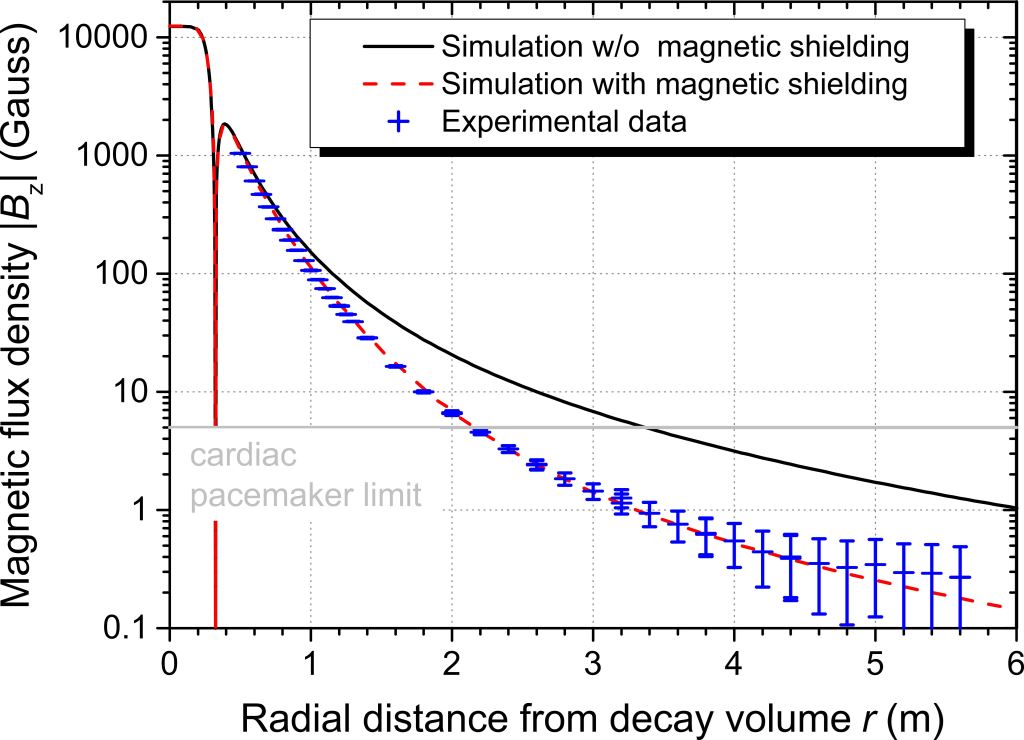}%
	\caption{Influence of the magnetic shielding shown in Fig.~\ref{fig:photo} on the field between two billets:  Magnetic flux density $B(r)$ in radial direction from the decay volume (DV) of the spectrometer \aspect, without (black) and with shielding (dashed red), for a working current of $I_{\rm main}=40$\,A.  The blue crosses show the results from a measurement at the TRIGA Mainz, in accordance with the simulation.  The cardiac pacemaker limit (gray line) of 5\,Gauss is reached at a radial distance of 2.2\,m from the DV.}%
	\label{fig:radialdata}%
\end{figure}

The simulation of the exterior magnetic field has been verified experimentally at the TRIGA Mainz with a Hall probe\footnote{A Group3 Technologies miniature hall probe MPT-141 \cite{MPT141}.}.   For this end, the exterior magnetic field has been measured for {\aspect} magnet turned 'on' and 'off', and the experimental data 'off' have been subtracted from 'on'.  As can been seen from Fig.~\ref{fig:radialdata},$^{\ref{foo:SimulatedData}}$ the stray magnetic field is suppressed to 0.35\,Gauss in a radial distance of 5\,m from the DV, which corresponds to 0.85\,Gauss ($<0.1$\,mT) for the design current of $I_{\rm main}=100$\,A, as was required.  Up to a distance of 4.8\,m, the experimental data points perfectly\footnote{\label{foo:Deltarho} The mean deviation of $(-1.7 \pm 4.8)$\,\% can partly derive from a positioning inaccuracy $\delta \rho$ of our Hall probe of 1 to 5\,cm.}~~\footnote{\label{foo:DeltaHall}Deviations $<0.2$\,Gauss can be related to the absolute accuracy of our Hall probe of $\pm$0.18\,Gauss in the 0.3\,T measurement range.} match the simulated magnetic field values:  The stray magnetic field is decreased by a shielding factor of up to 6.6.

\noindent
But from a distance of 5\,m, the experimental data deviate by 30 to 50\,\% from the simulated values.  We believe that the deviations are caused by other magnetic material.$^{\ref{foo:DeltaHall}}$  And unfortunately, due to lack in space, we have not yet been able to check the reduction of the exterior magnetic field by a shielding factor of 8.3 from a radial distance of 10\,m from the DV.

\begin{figure}[t]%
	\begin{minipage}[t]{0.475\textwidth}
  	\includegraphics[width=0.96\textwidth]{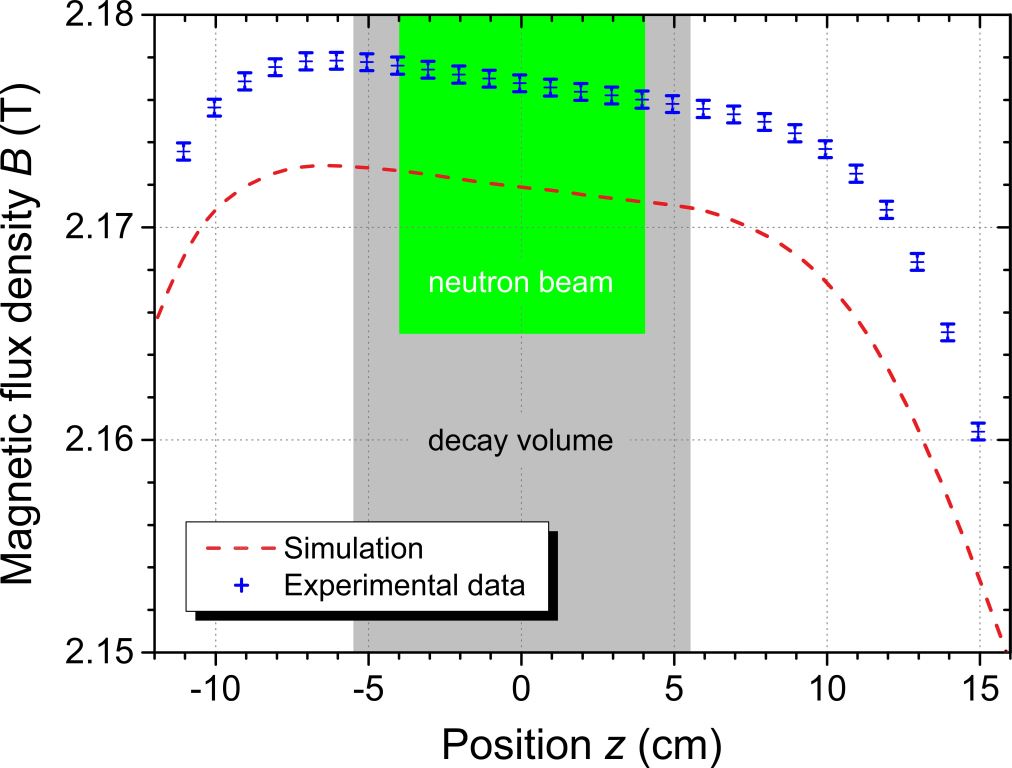}%
		\caption{Measured influence of the magnetic shielding shown in Fig.~\ref{fig:photo} on the field in the decay volume (gray bar) of \aspect:  The blue crosses show the results from a measurement of the magnetic flux density $B(z)$ along the symmetry axis of the magnet at the ILL, for working currents of $I_{\rm main}=70$\,A, $I_3=35$\,A, and $I_5=15$\,A.  The measured values exceed the simulated values (dashed red) by 0.2\,\%.}%
		\label{fig:DVdata}%
	\end{minipage}
	\hfill
	\begin{minipage}[t]{0.475\textwidth}
  	\includegraphics[width=\textwidth]{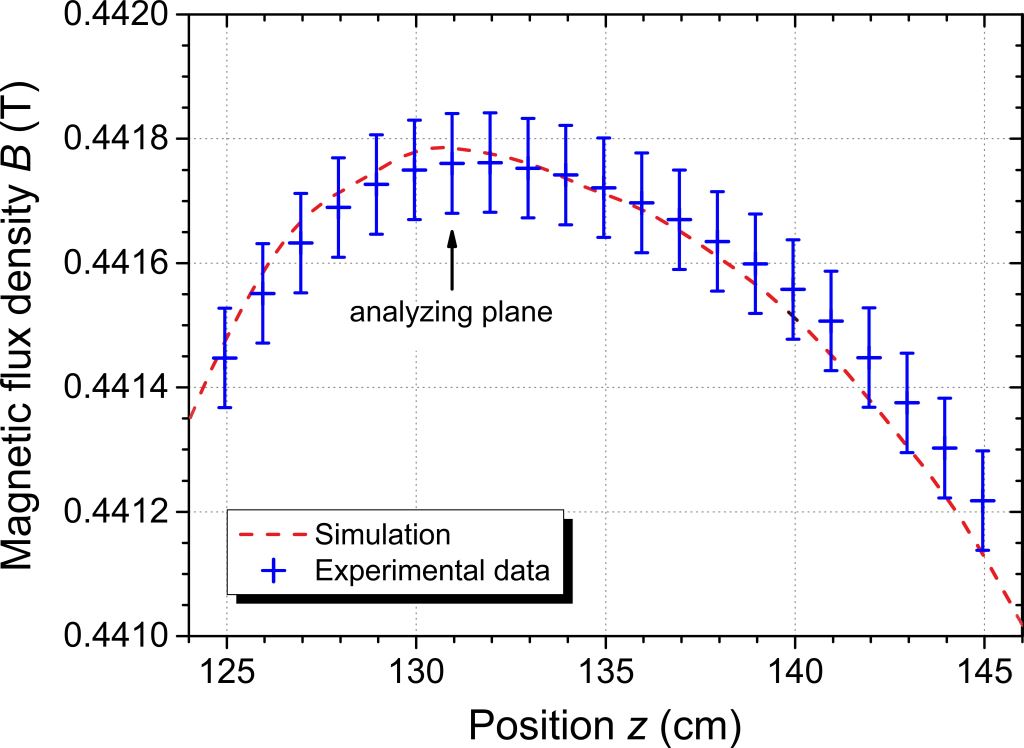}%
		\caption{Measured influence of the magnetic shielding shown in Fig.~\ref{fig:photo} on the field in the analyzing plane of the spectrometer \aspect:  The blue crosses show the results from a measurement of the magnetic flux density $B(z)$ along the symmetry axis of the magnet at the ILL, for a working current of $I_{\rm main}=70$\,A.  The
measured values perfectly (cf. Fn.~\ref{foo:Deltaz}) match the simulated values (dashed red).}%
		\label{fig:APdata}%
	\end{minipage}
\end{figure}

The last step to verify and validate the simulation model was to measure the internal magnetic field of \aspect.  There are two methods:  On-line with nuclear magnetic resonance (NMR) probes and off-line with a Hall probe.  The operating temperature range of our Hall probe is 0 to 50$^\circ$\,C \cite{MPT141}, while the {\aspect} magnet is operated at cryogenic temperatures.  For magnetic field measurements with the Hall probe, the electrodes schematically shown in Fig.~\ref{fig:Sketch} have been removed from the cold bore tube of the magnet and an inverted\footnote{\label{foo:Dewar}The dewar was built to maintain its surroundings ({\aspect} magnet) at cryogenic temperature, while its content (air) at ambient temperature.} non-magnetic dewar has to be inserted instead (see Ref.~\cite{ayala:2005} for details).  Then the magnetic field along and close to the symmetry axis of the magnet can be measured with our Hall probe, at about room temperature.  Figures~\ref{fig:DVdata} and~\ref{fig:APdata} show a measurement along the symmetry axis.  In the AP, the experimental data points match\footnote{\label{foo:Deltaz}A minor deviation of $<1\times10^{-4}$ can be attributed to a positioning inaccuracy $\delta z$ of our Hall probe of 5 to 10\,mm.} the simulated magnetic field values.  In the DV, the experimental data deviate by 0.2\,\% from the simulated values.  We believe that the deviations can be attributed to the moderate knowledge of the magnetization curves of soft iron and construction steel.  However, both the shape and the high homogeneity of the magnetic field in the DV remain unaffected by the return yoke, while the magnetic field ratio $r_{\rm B}$ (cf. Sec.~\ref{sec:aspect}) changes slightly by $-0.2$\,\%.  Again, this ratio change needs to be taken into account in the determination of the angular correlation coefficient $a$, but is otherwise irrelevant for the purpose of the {\aspect} experiment.

The accuracy achieved by NMR probes is superior to that obtained by Hall probes.  Therefore a NMR magnetometer was developed \cite{ayala:2011} to monitor on-line the magnetic field ratio $r_{\rm B}$.  For this purpose, two NMR probes are installed in the DV and in the AP, between the electrodes and the cold bore tube of the magnet (cf. Fig.~\ref{fig:Sketch}).  Measurements at the ILL have proven that the ratio is stable and reproducible within $2\times10^{-5}$ (relative) \cite{ayala:2011}.

\section{Conclusion and outlook}
\label{sec:conclusion}

The neutron decay spectrometer {\aspect} consists inter alia of a (superconducting) magnet system that generates a stray magnetic field of 4.5\,Gauss (0.4\,mT) in a radial distance of 5\,m from the decay volume (DV).  In order not to disturb other experiments in the vicinity of \aspect, we have designed, built, and tested successfully a magnetic field return yoke for the {\aspect} magnet system.  The FEA software package COMSOL Multiphysics$^{\textregistered}$ has been used to model and optimize its design:  A structure made of soft iron and construction steel serves to suppress the stray magnetic field to 0.65\,Gauss ($<0.1$\,mT) in a radial distance of 5\,m from the DV.

First the magnetic force calculations have been validated experimentally both at the TRIGA Mainz and at the ILL.  Then the simulation of the exterior magnetic field has been verified experimentally, up to a radial distance of 5.6\,m from the DV, at the TRIGA Mainz with a Hall probe.  And finally the simulations of the internal magnetic field have been checked experimentally at the ILL both with our Hall probe (at the beam position) and nuclear magnetic resonance probes:  Both the shape and the high homogeneity of the magnetic fields in the DV and the AP remain unaffected by the return yoke, while the ratio of the magnetic fields in the DV and the AP changes slightly.  The latter needs to be taken into account in the determination of the antineutrino-electron angular correlation coefficient $a$, but is otherwise irrelevant for the purpose of the {\aspect} experiment.

For the set-up of the new facility PERC at the FRM II one is confronted with a similar problem.  The main component of the instrument PERC is a more than 11\,m long superconducting magnet system, with a strong longitudinal magnetic field of up to 6\,T.  To comply with the FRM II safety regulations, the PERC magnet system will be surrounded with a structure that expands the return yoke presented here by additional steel plates.   COMSOL Multiphysics$^{\textregistered}$ has also been used to determine the most suitable geometry for the magnetic shielding for PERC.  But, unlike \aspect, the magnet geometry is non-axially symmetric and has therefore to be simulated fully in three dimensions.

\section*{Acknowledgements}
\label{sec:acknowledgements}

We appreciated the support of K. Eberhardt and the staff of the TRIGA Mainz.  We are grateful to L. Cabrera Brito, Ch. Palmer, K.\,K.\,H. Leung, M. Simson, and F. Chedane for their help in the magnetic field measurements.  We thank M. Hehn and the staff of the MPIP Mainz for their contributions to the electronics of our nuclear magnetic resonance probes.  We appreciated the support of T. Soldner, D. Jullien, and the staff of the ILL during the magnetic field measurements.  This work was supported by the German Federal Ministry for Research and Education under Contract No. 06MZ989I, 06MZ170, by the European Commission under Contract No. 506065, and by the Universit\"at Mainz.





\bibliographystyle{elsarticle-num}
\bibliography{Bibliography}

\begin{thebibliography}{10}
\expandafter\ifx\csname url\endcsname\relax
  \def\url#1{\texttt{#1}}\fi
\expandafter\ifx\csname urlprefix\endcsname\relax\def\urlprefix{URL }\fi
\expandafter\ifx\csname href\endcsname\relax
  \def\href#1#2{#2} \def\path#1{#1}\fi

\bibitem{zimmer:2000}
O.~Zimmer, et~al., ``{\aspect}'' – a new spectrometer for the measurement of
  the angular correlation coefficient $a$ in neutron beta decay, Nucl. Instr.
  and Meth. A 440 (2000) 548.

\bibitem{glueck:2005}
F.~Gl{\"u}ck, et~al., The neutron decay retardation spectrometer {\aspect}:
  Electromagnetic design and systematic effects, Eur. Phys. J. A 23 (2005) 135.

\bibitem{baessler:2008}
S.~Bae{\ss}ler, et~al., First measurements with the neutron decay spectrometer
  {\aspect}, Eur. Phys. J. A 38 (2008) 17.

\bibitem{PF1}
Polarised cold neutron facility {PF}1, in: {Y}ellow {B}ook (2008 ed.),
  {I}nstitut {L}aue-{L}angevin, {G}renoble, {F}rance, 2008, pp. 110$-$111,
  \url{http://www.ill.eu/fileadmin/users_files/Other_Sites/YellowBook2008CDRom/page/pg.htm?rub=9_1}.

\bibitem{konrad:2009}
G.~Konrad, et~al., {T}he {P}roton {S}pectrum in {N}eutron {B}eta {D}ecay:
  {L}atest {R}esults with the {\aspect} {S}pectrometer, Nucl. Phys. A 827
  (2009) 529c.

\bibitem{simson:2009}
M.~Simson, et~al., {M}easuring the proton spectrum in neutron decay - latest
  results with {\aspect}, Nucl. Instr. and Meth. A 611 (2009) 203, in
  \textit{Proceedings of the International Workshop on Particle Physics with
  Slow Neutrons, Grenoble, France, 2008}, \url{arXiv:0811.3851v1 [nucl-ex]}
  (2008).

\bibitem{IN11}
Spin-echo spectrometer {IN}11, in: {Y}ellow {B}ook (2008 ed.), {I}nstitut
  {L}aue-{L}angevin, {G}renoble, {F}rance, 2008, pp. 90$-$91,
  \url{http://www.ill.eu/fileadmin/users_files/Other_Sites/YellowBook2008CDRom/page/pg.htm?rub=7_4}.

\bibitem{konrad:2007}
G.~Konrad, et~al., {D}esign of an {A}nti-{M}agnetic {S}creen for the {N}eutron
  {D}ecay {S}pectrometer {\aspect}, in: Proceedings of the European Comsol
  Conference 2007, Grenoble, France, 2007, pp. 241--245,
  \url{http://www.comsol.com/papers/3276/}.

\bibitem{konrad:2011c}
G.~Konrad, Measurement of the {P}roton {R}ecoil {S}pectrum in {N}eutron {B}eta
  {D}ecay with the {S}pectrometer {\aspect}: {S}tudy of {S}ystematic {E}ffects,
  Ph.D. thesis, Johannes Gutenberg-Universit{\"a}t {M}ainz,
  \url{http://ubm.opus.hbz-nrw.de/volltexte/2012/3053/} (2011).

\bibitem{earnshaw:1842}
S.~Earnshaw, On the {N}ature of the {M}olecular {F}orces which {R}egulate the
  constitution of the {L}uminifereous {E}ther, Trans. Camb. Phil. Soc. 7 (1842)
  97--112.

\bibitem{abele:1997}
H.~Abele, et~al., A measurement of the beta asymmetry {$A$} in the decay of
  free neutrons, Phys. Lett. B 407 (1997) 212.

\bibitem{abele:2002}
H.~Abele, et~al., Is the unitarity of the quark-mixing {CKM} matrix violated in
  neutron beta-decay?, Phys. Rev. Lett. 88 (2002) 211801.

\bibitem{kreuz:2005a}
M.~Kreuz, et~al., A measurement of the antineutrino asymmetry {$B$} in free
  neutron decay, Phys. Lett. B 619 (2005) 263.

\bibitem{schumann:2007c}
M.~Schumann, et~al., Measurement of the {N}eutrino {A}symmetry {P}arameter
  {$B$} in {N}eutron {D}ecay, Phys. Rev. Lett. 99 (2007) 191803.

\bibitem{schumann:2008b}
M.~Schumann, et~al., Measurement of the {P}roton {A}symmetry {P}arameter {$C$}
  in {N}eutron {B}eta {D}ecay, Phys. Rev. Lett. 100 (2008) 151801,
  arXiv:0712.2442 [hep-ph].

\bibitem{mund:2013}
D.~Mund, et~al., Determination of the {W}eak {A}xial {V}ector {C}oupling from a
  {M}easurement of the {B}eta {A}symmetry {P}arameter {$A$} in {N}eutron {B}eta
  {D}ecay, Phys. Rev. Lett. 110 (2013) 172502, \url{arXiv:1204.0013 [hep-ex]}
  (2012).

\bibitem{maerkisch:2006}
B.~M\"arkisch, Das {S}pektrometer {PERKEO III} und der {Z}erfall des freien
  {N}eutrons, Ph.D. thesis, Ruperto-Carola University of Heidelberg (2006).

\bibitem{maerkisch:2009}
B.~M\"arkisch, et~al., The new neutron decay spectrometer {PERKEO III}, Nucl.
  Instr. and Meth. A 611 (2009) 216.

\bibitem{mest:2011}
H.~Mest, Measurement of the $\beta$-{A}symmetry in the {D}ecay of {F}ree
  {P}olarized {N}eutrons with the {S}pectrometer {PERKEO III}, Ph.D. thesis,
  Ruperto-Carola University of Heidelberg (2011).

\bibitem{wietfeldt:2005b}
F.~E. Wietfeldt, et~al., A method for an improved measurement of the
  electron-antineutrino correlation in free neutron beta decay, Nucl. Instr.
  and Meth. A 545 (2005) 181.

\bibitem{wietfeldt:2009}
F.~E. Wietfeldt, et~al., {aCORN}: An experiment to measure the
  electron-antineutrino correlation in neutron decay, Nucl. Instr. and Meth. A
  611 (2009) 207.

\bibitem{bowman:2005}
J.~Bowman, On the {M}easurement of the {E}lectron-{N}eutrino {C}orrelation in
  {N}eutron {B}eta {D}ecay, J. Res. Natl. Inst. Stand. Technol. 110 (2005) 407.

\bibitem{pocanic:2009}
D.~Po\v{c}ani$\acute{\rm c}$, et~al., Nab: Measurement principles, apparatus
  and uncertainties, Nucl. Instr. and Meth. A 611 (2009) 211.

\bibitem{alarcon:2010}
R.~Alarcon, et~al., Nab proposal update and funding request,
  \url{http://nab.phys.virginia.edu/nab\_doe\_fund\_prop.pdf} (2010).

\bibitem{baessler:2012}
S.~Bae{\ss}ler, et~al., Neutron {B}eta {D}ecay {S}tudies with {Nab},
  \url{arXiv:1209.4663v1 [nucl-ex]} (2012).

\bibitem{dubbers:2008}
D.~Dubbers, et~al., {A} clean, bright and versatile source of neutron decay
  products, Nucl. Instr. and Meth. A 596 (2008) 238, \\for an extended version,
  see \url{arXiv:0709.4440v1 [nucl-ex]} (2007).

\bibitem{konrad:2012}
G.~Konrad, et~al., Neutron {D}ecay with {PERC}: a {P}rogress {R}eport, J.
  Phys.: Conf. Series 340 (2012) 012048.

\bibitem{haiden:2013}
P.~Haiden, Design of the {M}agnetic {S}hielding for {PERC}, Master's thesis,
  Technische Universit\"at Wien (2013).

\bibitem{haiden:2014}
P.~Haiden, et~al., Design of the {M}agnetic {S}hielding for {PERC}, to be
  published.

\bibitem{beck:2003}
M.~Beck, et~al., {WITCH}: a recoil spectrometer for weak interaction and
  nuclear physics studies, Nucl. Instr. and Meth. A 503 (2003) 567.

\bibitem{beck:2011a}
M.~Beck, et~al., First detection and energy measurement of recoil ions
  following beta decay in a penning trap with the {WITCH} experiment, Eur.
  Phys. J. A 47 (2011) 45.

\bibitem{habs:2000}
D.~Habs, et~al., The {REX-ISOLDE} project, Hyperfine Interact. 129 (2000) 43.

\bibitem{REXISOLDE}
Radioactive ion beam facility {REX-ISOLDE},
  \url{http://isolde.web.cern.ch/isolde/REX-ISOLDE/}.

\bibitem{tandecki:2011}
M.~Tandecki, Progress at the {WITCH} {E}xperiment towards {W}eak {I}nteraction
  {S}tudies, Ph.D. thesis, Kath. Univ. Leuven (2011).

\bibitem{garrett:1963}
M.~W. Garrett, Calculation of fields, forces, and mutual inductances of current
  systems by elliptic integrals, J. Appl. Phys. 34 (1963) 2567.

\bibitem{TOSCA}
{TOSCA} analysis package,
  \url{http://www.chilton-computing.org.uk/inf/eng/electromagnetics/p001.htm}.

\bibitem{COMSOL}
Finite element analysis software package {COMSOL} {M}ultiphysics,
  \url{http://www.comsol.com/}.

\bibitem{ayala:2005}
F.~Ayala~Guardia, {F}irst {T}ests of the neutron decay spectrometer {\aspect},
  Master's thesis, Johannes Gutenberg-Universit{\"a}t {M}ainz (2005).

\bibitem{ayala:2011}
F.~Ayala~Guardia, Calibration of the retardation spectrometer {\aspect}, Ph.D.
  thesis, Johannes Gutenberg-Universit{\"a}t {M}ainz,
  \\\url{http://ubm.opus.hbz-nrw.de/volltexte/2012/3012/} (2011).

\bibitem{babic:2000}
S.~Babic, C.~Akyel, M.~M. Gavrilovic, Calculation improvement of 3d linear
  magnetostatic field based on fictitious magnetic surface charge, IEEE Trans.
  Magn. 36 (2000) 3125.
\newblock \href {http://dx.doi.org/10.1109/20.908707}
  {\path{doi:10.1109/20.908707}}.

\bibitem{glueck:2011}
F.~Gl{\"u}ck, {A}xisymmetric magnetic field calculation with zonal harmonic
  expansion, PIER B 32 (2011) 351,
  \\\url{http://www.jpier.org/pierb/pier.php?paper=11042108}.

\bibitem{bozorth:1961}
R.~M. Bozorth, Ferromagnetism, Wiley, 1961.

\bibitem{boll:1990}
R.~Boll, {W}eichmagnetische {W}erkstoffe: {E}inf\"uhrung in den {M}agnetismus.
  {VAC}-{W}erkstoffe und ihre {A}nwendung, 4th Edition, Publicis Corporate
  Publishing, 1990.

\bibitem{lederer:1998}
D.~Lederer, A.~Kost, Modelling of {N}onlinear {M}agnetic {M}aterial {U}sing a
  {C}omplex {E}ffective {R}eluctivity, IEEE Trans. Magn. 34 (1998) 3060.

\bibitem{brammer:1995}
U.~Brammer, {S}chirmung supraleitender magnetischer {E}nergiespeicher, Ph.D.
  thesis, Technische Universit\"at M\"unchen (1995).

\bibitem{ludwig:2006}
U.~Ludwig-Mertin, private communication (2006).

\bibitem{spielvogel:2006}
O.~Spielvogel, private communication (2006).

\bibitem{MPT141}
{Group3 Technology Ltd.},
  \url{http://www.westlakemeter.com/Product/Group/MPT-141\%20spec.pdf}.

\bibitem{wunderle:2012}
A.~Wunderle, private communication (2012).

\end{thebibliography}







\end{document}